\documentclass[twoside]{article}

\usepackage[preprint]{aistats2026}
\usepackage{subcaption}
\usepackage[round]{natbib}

\usepackage[utf8]{inputenc} 
\usepackage[T1]{fontenc}    
\usepackage{hyperref}       
\usepackage{url}            
\usepackage{booktabs}       
\usepackage{amsfonts}       
\usepackage{nicefrac}       
\usepackage{microtype}      
\usepackage{xcolor}         
\usepackage{graphicx}
\usepackage{floatrow}
\usepackage{comment}
\usepackage{multirow}
\usepackage{amsmath}
\usepackage{amssymb}
\usepackage{xcolor}
\usepackage{lscape}
\usepackage{adjustbox}

\begin{document}

%
\runningtitle{Life Sequence Transformer}

%

\twocolumn[

\aistatstitle{Life Sequence Transformer: Generative Modelling of Socio-Economic Trajectories from Administrative Data}

\aistatsauthor{ Alberto Cabezas \And Carlotta Montorsi}

\aistatsaddress{ Department of Economics and Statistics \\
University of Turin  \And
Luxembourg Institute of \\ Socio-Economic Research } ]

\begin{abstract}
Generative modelling with Transformer architectures can simulate complex sequential structures across various 
applications. We extend this line of work to the social sciences by introducing a Transformer-based generative model tailored to longitudinal socio-economic data. Our contributions are: (i) we design a novel encoding method that represents socio-economic life histories as sequences, including overlapping events across life domains; and (ii) we adapt generative modelling techniques to simulate plausible alternative life trajectories conditioned on past histories. Using large-scale data from the Italian social security administration (INPS), we show that the model can be trained at scale, reproduces realistic labour market patterns consistent with known causal relationships, and generates coherent hypothetical life paths. 
This work demonstrates the feasibility of generative modelling for socio-economic trajectories and opens new opportunities for policy-oriented research, with counterfactual generation as a particularly promising application.
\end{abstract}

\section{INTRODUCTION}

In fields such as physics \citep{geneva2022transformers}, chemistry \citep{kim2021generative}, and health \citep{yang2023transformehr}, researchers successfully use Transformer-based generative models to simulate surrogate dynamical systems under varying conditions and estimate event probabilities. In addition to their predictive capabilities, generative models have the potential to explore hypothetical system evolutions under different scenarios.

We extend this line of research to the social sciences by proposing a Transformer-based generative model for sequences of human life events. Longitudinal administrative records contain rich information on individuals’ trajectories across labour markets, health, education, and other socio-economic domains, yet these data are rarely treated as sequences for generative modelling. Through a novel encoding method of socio-economic events, we enable the training of a decoder-only Transformer model capable of generating plausible synthetic life trajectories.

Our model is trained on 1.6 million individual labour market histories from the Italian social security administration (INPS), described in Section \ref{data_description}. The representation of these histories as overlapping socio-economic events encoded into life sequences is introduced in Section \ref{data_encoding}. The architecture and training procedure of our life sequence Transformer are detailed in Section \ref{model}. The empirical analysis in Section \ref{empirical} illustrates the generative potential of the framework through two benchmarking exercises: the accuracy of simulated sequences in capturing established causal nexus among life events (Section \ref{applications}) and the predictive accuracy on out-of-sample trajectories (Section \ref{out_of_sample}).

Once trained, the model can generate realistic continuations of partially input sequences and simulate alternative life paths. In other words, it can create counterfactuals, i.e., plausible ``what if'' scenarios that describe how events might have unfolded under different circumstances. This ability provides a scalable framework for constructing coherent counterfactual trajectories without the need for external control groups. In Section \ref{discussion}, we elaborate on how this approach can be integrated into experimental designs in the social sciences and outline promising directions for future research that build on our framework.

\section{BACKGROUND}

Our work builds on recent advances in Transformer-based sequence modelling for longitudinal labour and health data. In particularly, Life2Vec \citep{savcisens2024using}, CAREER \citep{vafa2024careerfoundationmodellabor}, and TransformEHR \citep{yang2023transformehr}. Life2Vec tests the validity of encoder-only Transformer models for modelling socio-economic administrative data from Denmark. It outperforms the predictive performance of standard models, such as mortality tables, logistic regression, or recurrent neural networks, in predicting an individual's future life outcomes (e.g., early mortality) or time-invariant characteristics (e.g., personality nuances). However, this model lacks generative capability to simulate sequences of future tokens, and the data encoding cannot be easily adapted to a generative task. In contrast, our proposed encoding approach enables the training of a generative decoder, allowing for the simulation of counterfactual life sequences on a model architecture directly inspired by Life2Vec.

The CAREER encoder-decoder Transformer model focuses on job sequence generation with annual temporal resolution and no rich occupational attributes, such as firm size, social security benefits, or multiple job episodes within the same year. We enhance this framework by introducing fine-grained temporal encoding and a richer set of characteristics that span labour and demographic dimensions. In CAREER, time is introduced with a timestep of one year, and there is no way to model the duration of events or the exact timing of their occurrence. You could increase the timestep to monthly steps, but this would restrict generation to a fixed monthly grid, offering no control over when events begin or how long they last within the year. Our method of encoding time, by contrast, enables flexible temporal generation by allowing each life event to have a defined starting month and duration in months within the year.

TransformEHR is an encoder-decoder Transformer generative model that encodes sequences of electronic health records and predicts the subsequent closest diagnoses (ICD-10 codes). It includes demographic and time embeddings to model the numerical format of the visit date. TransformEHR is pre-trained to predict a patient's future visits with ICD codes, given longitudinal codes up to the current visit. Similarly to Life2Vec, the timing of each visit must be specified at inference time, meaning the model cannot decide when events occur, but only predict diagnoses for predefined dates. In contrast, our model allows the generation process to determine when events occur and their respective durations. Crucially, our model supports variable event durations, which is essential for modelling life sequences that extend beyond instantaneous diagnoses.



\section{WHIP DATA} \label{data_description}

Our model is trained on the Work History Italian Panel (WHIP), a 1:15 sample of individuals drawn from the INPS database \citep{leombruni2010note, bena2012new}. INPS records cover most private-sector employees, quasi-dependent and self-employed workers, as well as periods of retirement and access to social benefits (e.g., unemployment or maternity leave). The WHIP provides detailed longitudinal employment and benefit histories for over 4 million individuals.

We use the WHIP database for the period 1990–2015 and apply sample selection criteria to obtain sequences that are sufficiently long to provide meaningful information. Specifically, we retain only individuals with complete information up to 2015, at least half of their potential labour force participation period observed, non-missing birth dates, at least five records in total, plausible entry ages (between 15 and 80), and a labour force attachment of at least 10\%. This resulted in a sample of around 1.65 million life trajectories. A full description of the dataset and details of the selection procedure are provided in Appendix~\ref{whip_data}.

Each labour event contains timing (start month and duration) and income, which is standardised into monthly real terms and discretised into 100 quantiles. Yearly income is the principal variable for the social security system, as it forms the basis for calculating contributions and benefits. Each income record is linked to a range of demographic, workplace, and labour-force-specific attributes. Demographic information includes sex, birth year, and month, and the macro-area of birth. Workplace characteristics cover the sector of activity, enterprise size, and workplace location at the provincial level.

Labour force information is recorded in terms of labour status (e.g., employee, self-employed, pensioner, unemployed) with additional modifiers depending on status. For employees, these include job title, work intensity, number of weeks of maternity leave, number of weeks of sick leave, and whether the contract is part-time or full-time. Together, these attributes enable a detailed reconstruction of individuals’ trajectories across work, unemployment, pensions, and benefit receipt. A full description of all 15 variables and encoding details is provided in Appendix~\ref{whip_labour}.

\section{DATA ENCODING} \label{data_encoding}

\begin{figure*}[!thp]
\centering
\includegraphics[width=1\columnwidth]{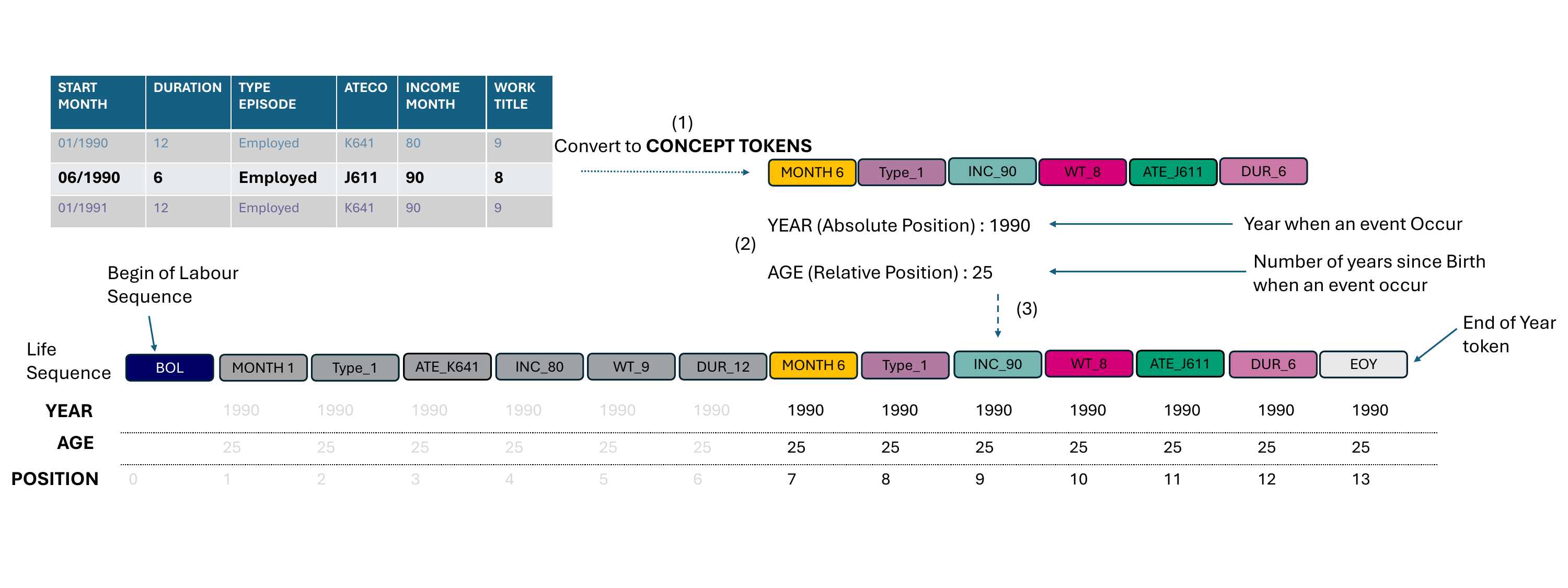}%
\caption{The steps for converting data from Tabular format to the Generative model suitable format: (1) Convert relevant features to concept tokens (2) Look up relevant temporal information such as age and year (3) Concatenate token, time, and positional embeddings to the sequence.}
\label{fig:tabular2token}
\end{figure*}

We build on the work of \cite{savcisens2024using}, where life events are concatenated in a sequential structure, with absolute and relative temporal information embedded as auxiliary attributes rather than forming the structural backbone of the sequence. Instead, our encoding method imposes an explicit calendar-like structure on each life trajectory to support the sequential generation of future tokens. The entire observed work history is divided into yearly segments, each further subdivided into twelve monthly intervals, which are then populated with event information as needed. Life events are represented as discrete units with attributes such as employment type, income quantiles, work intensity, sector, qualification, and duration, forming a sequence of socio-economic states. 

For instance, if an individual remains employed at the same firm for 10 uninterrupted years, we encode this as 10 events, each lasting 12 months, described using the same set of descriptive tokens, and with a likely increase in income or a change in work title if promoted over the years. 
If a person holds a one-year contract but experiences a one-month interruption, the year is split into two separate events with different durations. Though the same set of concept-tokens is otherwise used to describe these events. When no event occurs within a year, the structure still records an ``empty'' year 
preserving the regular yearly pace of the sequence. 

Figure~\ref{fig:tabular2token} illustrates the encoding pipeline. We consider each tabular record (i.e., each row in the administrative dataset) as an \textit{event} and translate each column describing it into a \textit{concept-token}. Then, we concatenate all the events for a given individual in chronological order and create an individual sequence. 

Each event includes embeddings on positional information: year or \textit{absolute} position reporting the calendar year when the event occurred, age or \textit{relative} position with the individual’s age at the time of the event, and the usual \textit{positional} embedding used in Transformer models for the position of the token in the sequence. 
Notice, these three embeddings are not part of the generative task; instead, they are deterministic. 

Sequences start with the special token \texttt{<BOL>} (Beginning of Life). Individual events are then encoded using a fixed token order. \texttt{<MONTH\_X>} marks the calendar month the event begins and helps separate multiple events within the same year. Next comes \texttt{<TYPE\_X>} to indicate the event type (e.g., employment, unemployment), optionally followed by modifiers like \texttt{<ATECO\_X>} (for sector) and \texttt{<WRK\_TITLE\_X>} (for job title), depending on the event. \texttt{<DUR\_X>} defines the event’s duration in months and concludes the block describing the event. If more events follow in the same year, another \texttt{<MONTH\_X>} appears; otherwise, the \texttt{<EOY>} (End of Year) token concludes the description of the events in a year. If no further events ensue, the entire life sequence ends with an \texttt{<EOL>} (End of Life) token. Each sequence maintains a consistent ordering of these concept-tokens, ensuring a structured representation and facilitating the learning of this artificial language.

While the experiments in this paper focus on labour-related life sequences, we emphasise that the proposed encoding scheme applies to other life domains, such as health and education. In the health domain, for instance, life sequences could represent concatenations of diagnostic attributes, annotated with the month of diagnosis and the duration of treatment or hospitalisation. Although a monthly resolution may appear too coarse for modelling certain health events, this limitation is mitigated by including intensity tokens \texttt{<INTENSITY\_X>}, which capture the extent of engagement within each event's duration. For a detailed description of the data and a labour life sequence example, see Appendix \ref{whip_data}.

Background time-invariant information, such as gender, place, and date of birth, is also represented using concept-tokens (e.g., \texttt{<F>} or \texttt{<M>} for female and male). This information is placed at the beginning of the sequence, forming the entire input as a single stream of tokens. The encoding distinguishes between static demographic attributes and the temporal sequence of life events, which begins with the \texttt{<BOL>} token.



\section{TRANSFORMER MODEL} \label{model}


We propose a decoder-only Transformer architecture, see Figure \ref{fig: decoder only}, trained autoregressively to predict the next token in the sequence, conditioned on past events and background information.

\subsection{Model Architecture}

The model follows a Transformer decoder-only design, consisting of three components: an embedding block, stacked Transformer-based decoder layers, and a task-specific output head. We make three key modifications to the standard architecture. First, we replace the conventional attention mechanism with Performer Attention \citep{choromanski2022rethinkingattentionperformers}, which reduces the number of trainable parameters and improves computational efficiency. Second, we incorporate Rotary Position Embeddings (RoPE) \citep{su2023roformerenhancedtransformerrotary}, adapted to work with Performer Attention, to better capture temporal patterns in the sequence. Third, to stabilise training, we apply RMS Normalisation \citep{zhang2019root} before the attention and feedforward layers. The decoder uses causal (or masked) self-attention, ensuring that each token can only attend to earlier positions in the sequence. 
During training, the model is optimised using a next-token prediction objective, learning to generate coherent event sequences conditioned on prior events and time-invariant variables.


\paragraph{Token embeddings} 

The first transformation of the input sequence occurs in the embedding layer, where each token in the sequence $S_r$ is mapped to a continuous vector representation using an embedding matrix $\mathcal{E}_\mathcal{V}: \mathcal{V} \rightarrow \mathbb{R}^d$. Each row of $\mathcal{E}_\mathcal{V}$ corresponds to a unique token from the vocabulary $\mathcal{V}$, and each column represents one of the $d$ embedding hidden dimensions. Following \citet{savcisens2024using}, we center the embeddings by removing the mean from each matrix row.

\paragraph{Time embeddings}
To encode temporal information, i.e., age and calendar year, we use the Time2Vec method \citep{kazemi2019time2veclearningvectorrepresentation}, which allows the model to capture both linear and periodic aspects of time. This method uses trainable parameters $\omega$ and $\varphi$ to define a flexible sinusoidal function. We construct two separate time embeddings: one for age, denoted by $\mathcal{T}_A: \mathcal{A} \rightarrow \mathbb{R}^d$ with $Z := 4$, and one for year, $\mathcal{T}_Y: \mathcal{Y} \rightarrow \mathbb{R}^d$ with $Z := 2$. These embeddings are computed as:

\begin{equation}
\mathcal{T}(x)[z] = \begin{cases}
\tanh(w_z x + \varphi_z) & \text{if } z \in [0, d//Z)\\
\sin(w_z x + \varphi_z) & \text{if } z \in [d//Z , d)
\end{cases}
\end{equation}

Initial iterations of the model used standard Time2Vec encoding, i.e., a linear component with trigonometric functions. However, we noticed that it led to poor out-of-sample performance: the model quickly lost coherence and failed to maintain the grammar of the life sequence when extrapolating beyond the training period. We hypothesised that this issue stemmed from the unbounded linear trend in the Time2Vec representation of age and year. Additionally, trigonometric components appeared to mimic linear trends, particularly for the year variable, and also for age. 

Hence, we adapt the original Time2Vec by varying the proportion of linear and periodic components for different temporal signals and bounding the linear component for better out-of-sample generalisation (see Section \ref{out_of_sample}). For year embeddings, we allocate a larger share of the hidden dimensions to the linear component ($\tanh$) to better capture the short-range time frame, which spans only 25 years. For the Age embedding, which spans 70 years in our dataset, we expand periodicity ($\sin$) to reflect cyclical patterns (e.g., retirement or career phases) that recur across broader age bands. 

\paragraph{Positional embeddings} In addition to temporal embeddings, we incorporate standard positional encodings to capture the order of tokens within each sequence. Our approach is adapted to use with local and global heads of Performer attention \citep{su2023roformerenhancedtransformerrotary}: sinusoidal positional encodings combined with RoPE in the global attention heads. Both positional and time embeddings are added to the token embeddings through a ReZero gate \citep{bachlechner2021rezero} before being passed to the attention layers.




\subsection{Training Procedure} \label{training}

\begin{table*}[!htbp]
\centering
\caption{Test set metrics for next-token prediction, macro-averaged across the vocabulary, conditioning on 0, 5, 1 and 10 initial years of sequence.}

\begin{tabular}{lccccc}
\hline
 & \textbf{Accuracy} & \textbf{Precision} & \textbf{Recall} & \textbf{F1 Score} & \textbf{Perplexity}\\
\hline
0 known years & 0.6225 & 0.6868 & 0.6225 & 0.6427 & 0.7288 \\
1 known year & 0.6773 & 0.7216 & 0.6773 & 0.6895 & 0.6937 \\
5 known years & 0.6883 & 0.7285 & 0.6883 & 0.6991 & 0.6727 \\
10 known years & 0.7039 & 0.7409 & 0.7039 & 0.713 & 0.6466 \\
\hline
\end{tabular}
\label{tab:next_token_metrics_decoder}
\end{table*}

Our dataset consists of 1,652,877 individual life sequences. We divide them into training (70\%), validation (15\%), and test (15\%) subsets. The models are trained for 15 to 20 epochs with early stopping. Each training batch contains 18 sequences, and gradients are accumulated over five steps before an optimiser step, resulting in an effective batch size of 90. 

\paragraph{Optimization}

We employ the \texttt{AdamW} optimiser \citep{loshchilov2019decoupledweightdecayregularization}, with no weight decay applied to bias terms, embedding weights, or normalisation layers. The learning rate is controlled by the One Cycle policy \citep{smith2019super} with a warm-up phase covering 30\% of the training steps. Gradient norms are clipped at a maximum value of 5 to prevent exploding gradients. To avoid overfitting, a dropout rate of 10\% is applied to the attention, feedforward, and embedding layers. Dataloaders and projection matrices of Performer attention are reloaded every epoch to ensure robustness in sampling. We use mixed-precision with \texttt{bfloat16}, which we chose to reduce memory usage and improve training efficiency, considering that RMS Normalisation has been shown to improve numerical stability in lower-precision regimes \citep{zhang2019root}.


\paragraph{Hyperparameters tuning}
We sample hyperparameters using the Tree-Structured Parzen Estimator \citep{watanabe2023tree} and Optuna \citep{optuna_2019}. Experiments showed that deeper models performed better. Without normalisation, the model misuses the embedding space, often requiring unusually large embedding dimensions to compensate. With RMS normalisation, this effect was mitigated, although the model benefited from longer warmup phases compared to setups using only ReZero. Increasing depth proved more effective than scaling up feedforward or embedding dimensions, or the number of random features. For local attention, we set the window size to cover approximately three complete events, which empirically balanced context coverage and memory use. Ultimately, we selected hyperparameters that maximized model capacity while fitting within the 16GB of VRAM available on an RTX A4000 GPU: 
10 decoder layers, with 8 attention heads, 6 local with a window size of 36 and 2 global with projection of 256 random features; embedding size of 240, and position-wise feed forward size of 960 on all layers. The model has around 10.1 million tunable parameters.

\section{EMPIRICAL ANALYSIS} \label{empirical}

In this section, we present the next-token prediction task metrics and evaluate the model’s ability to forecast out-of-sample for years beyond the training period. We then evaluate performance through a series of causal inference benchmarking exercises by comparing the generated life sequences against well-established empirical causal findings from labour economics. All experiments are run 
on the test subset of our data and code can be found in \href{https://github.com/albcab/life-sequence-transformer}{github/life-sequence-transformer}. 

\subsection{Next-Token Prediction Metrics} \label{evidence}

We begin by evaluating the model on a next-token prediction task over the test subset, which measures how accurately it generates the next element in a life sequence given prior tokens and background information.
Training each model took approximately 9 hours per epoch for the training phase and 1.5 hours per epoch for validation, totalling 10.5 hours per epoch. The final model was trained for 20 epochs for a total training time of 210 hours on a single NVIDIA RTX A4000 GPU with 16GB of VRAM.

Table \ref{tab:next_token_metrics_decoder} reports a standard set of classification metrics computed across the entire sequence dimension and vocabulary, excluding padding tokens. Each row corresponds to a different length of observed history provided as input. Specifically, we generate the continuation of a life sequence after observing either zero, one, five, or ten years of past events. The metrics include macro-averaged accuracy, precision, recall, and F1 score, each evaluated using top-1 predictions, i.e., consider only the highest probability to assign the predicted token. We also report the square root cross-entropy loss, i.e., perplexity, a widely used measure in language modelling.

As expected, the more historical context the model observes, the better it can predict future tokens. This is consistent with the intuition that individuals often follow relatively stable life patterns after a certain age, e.g., remain within certain industries over time. Access to a longer sequence of past events enables the model to learn person-specific trajectories and reduces uncertainty about what is likely to happen next. 

\subsection{Causal Benchmarks} \label{applications}

We evaluate the model’s capacity for generating causal relationships by testing whether it reproduces well-established effects from the labour economics literature and national policy. To do so, we truncate observed individual life sequences at specific cutoff years and condition the model on this truncated history. The model is then tasked with generating the continuation of the sequence forward in time. We only truncate at the level of whole years, so that the conditioning context always consists of a realistic, complete history up to the cutoff point. In each experiment, we choose the cutoff year to coincide with the onset of the event of interest (i.e., unemployment, maternity, or retirement), so that the subsequently generated trajectory can be compared to known empirical effects. Additional details of this procedure are provided in Appendix \ref{experimental_details}. All generation experiments are run on batches of 8 sequences. On a GPU, generating the next token takes approximately 0.2 seconds per batch, while on a CPU it takes about 2 seconds. 

\begin{table*}[!htbp]
\centering
\begin{tabular}{lccccr}
\hline
 & $\quad$ & $\text{ATE}_{\text{emp}}$ & $\text{ATE}_{\text{model}}$ & $\Delta$ & $n$ \\
\hline
Pension (1y cutoff)    &  &  7.47 [2.50, 12.23]  &  9.10 [4.95, 13.26]  &  1.62 [-0.36,  3.67]  & 1165 \\
Pension (4y cutoff)    &  &  7.47 [2.42, 12.59]  & 11.02 [7.07, 15.14]  &  3.55 [ 2.09,  5.07]  & 1165 \\
Unemployment $\pm$12m  &  &  4.18 [1.46,  6.67]  &  0.21 [-2.17,  2.55] & -3.98 [-6.38, -1.66]  &  153 \\
Unemployment $\pm$48m  &  &  3.41 [1.76,  4.90]  &  2.49 [1.08,  3.89]  & -0.92 [-2.18,  0.54]  &  556 \\
Unemployment $\pm$96m  &  &  7.25 [5.70,  8.70]  &  7.09 [5.86,  8.39]  & -0.16 [-1.36,  0.97]  & 1168 \\
Unemployment $\pm$144m &  &  8.27 [7.02,  9.51]  &  8.92 [7.76, 10.02]  &  0.65 [-0.34,  1.67]  & 1502 \\
Maternity 0--3 years (\%)   &  & -25.1 [-28.3, -22.2] & -22.6 [-24.6, -20.5] &  2.45 [-1.20,  6.20] &  631 \\
Maternity 0--5 years (\%)   &  & -19.1 [-22.0, -16.5] & -18.8 [-20.7, -16.8] &  0.31 [-2.90,  3.80] &  631 \\
Maternity 0--10 years (\%)  &  & -11.0 [-13.7,  -8.30] & -13.4 [-15.6, -11.2] & -2.35 [-5.90,  1.10] &  631 \\

\hline
\end{tabular}
\caption{Empirical and model-based average treatment effects (ATEs) point estimates with 95\% confidence interval in brackets. $\Delta$ is the signed difference between model and empirical ATEs. $n$ denotes the number of individuals in the analysis sample. For details on how ATEs are constructed in each experiment, see Appendix~\ref{experimental_details}.}
\label{ate}
\end{table*}

First, we evaluate whether the model captures the impact of an unemployment benefit policy rule in Italy. Until 2017, Italy's mobility allowance granted income support to workers displaced due to firm closure, with benefit duration varying discontinuously by age (cutoffs at 40 and 49) and by region (South vs. North/Central) \citep{brunello1997benefit}. For each individual, we truncate the sequence at the year when displacement is first observed and provide this truncated history to the model. The model must then generate forward, including the timing and duration of subsequent unemployment and benefit receipt.  

\begin{figure}[!thp]
\includegraphics[clip, trim=0.0cm 0.0cm 0.0cm 0.8cm, width=\columnwidth]{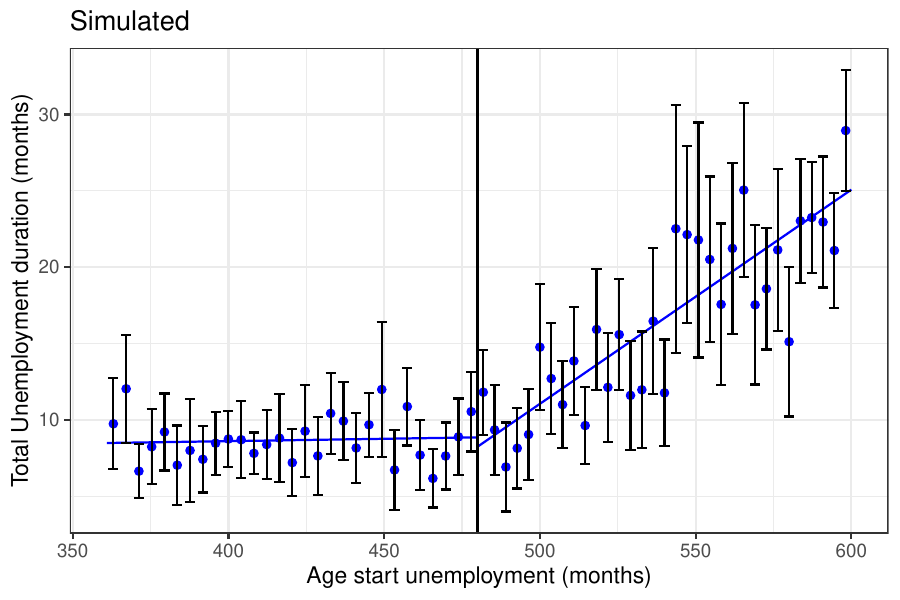}
\caption{Model-generated average duration with 90\% confidence interval of mobility allowance receipt duration as a function of worker age at job loss in months in a 10-year bandwidth around the 40-year-old threshold.}
\label{fig:unemployment_result_decoder_only}
\end{figure}

Figure \ref{fig:unemployment_result_decoder_only} shows the simulated average duration of mobility allowance receipt as a function of age at displacement. The simulated data captures the overall upward trend in benefit duration with age, suggesting that the model has learned the general relationship between age and benefit duration. However, it falls short of reproducing the sharp discontinuity observed at the threshold, instead smoothing over the abrupt policy-induced change. One likely reason is data scarcity: fewer than 1\% of sequences involve a mobility allowance spell (see Figure \ref{fig:labour_force_status_composition}), which restricts the amount of policy-relevant signal available for training.  

A more formal discontinuity analysis confirms this pattern. Table~\ref{ate} reports local average treatment effects (ATEs) for different bandwidths around the 40-year cutoff. At narrow bandwidths (e.g.,\ $\pm 12$ months), the model ATE fails to significantly capture the discontinuity. As the window expands to $\pm 48$ and $\pm 96$ months, however, the model’s ATEs move closer to the empirical benchmark, with discrepancies shrinking to below one month and confidence intervals overlapping zero. At the widest bandwidth ($\pm 144$ months), both the empirical and model estimates are around 8–9 months, with a modest and statistically insignificant difference.

The second exercise examines the motherhood penalty, i.e.,\ the long-term career setbacks women face after having children, often including reduced earnings and labour market detachment \citep{casarico2023behind, kleven2019children}. Here, we simulate women’s careers from the moment of first maternity. Specifically, we truncate sequences either (i) at the year of the first observed maternity leave (offset--0), or (ii) one year before that event (offset--1). The latter setup allows maternity leave itself to be generated or not, and enables counterfactual analysis by comparing career paths with and without maternity.

\begin{figure}[!htbp]
\centering
\includegraphics[width=\columnwidth]{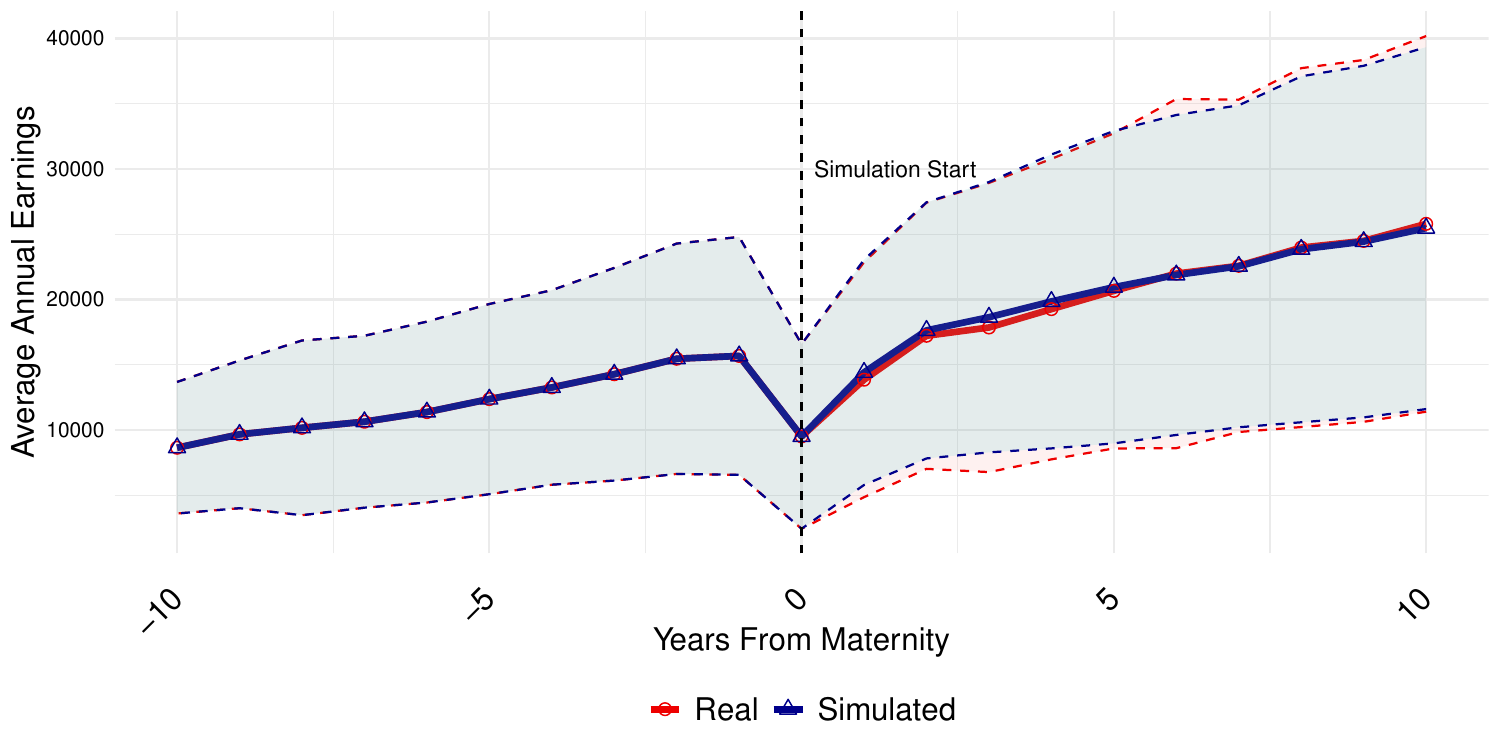}%
\caption{Real and model-generated trajectories of average annual earnings for working mothers in a 10-year window before and after their first maternity leave episode (offset--0). The solid lines represent average annual earnings, while the shaded areas denote the one standard deviation range.}
\label{fig:mother_result_decoder_only}
\end{figure}

Figure \ref{fig:mother_result_decoder_only} compares the real and simulated income trajectories of mothers following their first childbirth (see a breakdown by sector in Figure \ref{fig:generated_mother_ATECO}). Synthetic sequences are generated conditional on the real sequence up to the maternity leave year. The model successfully captures the key features of the observed trends, closely replicating the average post-birth income dynamics. 

To estimate the treatment effect of childbirth on earnings, we use an event-study design with controls for age and calendar year, as in \cite{kleven2019children}. Because fertility is endogenous, we construct counterfactual trajectories using both propensity-score matched non-mothers and model-generated no-maternity paths (offset--1 simulations). The details of this methodology are presented in Appendix~\ref{mothers}. These event-study coefficients are then summarised into average percentage income penalties over post-birth horizons, making them directly comparable to the ATEs reported for pension and unemployment.

The results show that the model reproduces the empirical child penalties closely across all horizons (Table~\ref{ate}). What is especially interesting is the pattern of differences: the model tends to understate the immediate short-run penalty but generates a more persistent effect in the medium and long run. In other words, the simulated trajectories suggest a smaller initial income drop but slower recovery, implying that the penalty lingers for longer. 

Figure \ref{fig:mother_result_synth} illustrates how the model-generated simulated controls (non-maternity paths) compare to real mothers’ earnings trajectories, with propensity score–matched mothers included as a comparison. The simulated controls track the counterfactual trend, reinforcing the finding that the model not only captures the observed income losses of mothers but also produces realistic no-maternity baselines.


The final exercise investigates whether the model replicates the well-documented effect of birth month on retirement timing \citep{ardito2020work}. Institutional features, such as school starting dates and labour market entry, generate systematic differences in retirement timing between individuals born in January versus those born in December. To test this, we truncate observed sequences prior to retirement. In one variant, we cut off at the year immediately before the observed retirement (so the model must decide whether retirement occurs in the following year or later). In another variant, we truncate four years before retirement, which increases the uncertainty in retirement timing and allows us to compare under different predictive horizons.

\begin{figure}[!htbp]
\centering
\includegraphics[width=\columnwidth]{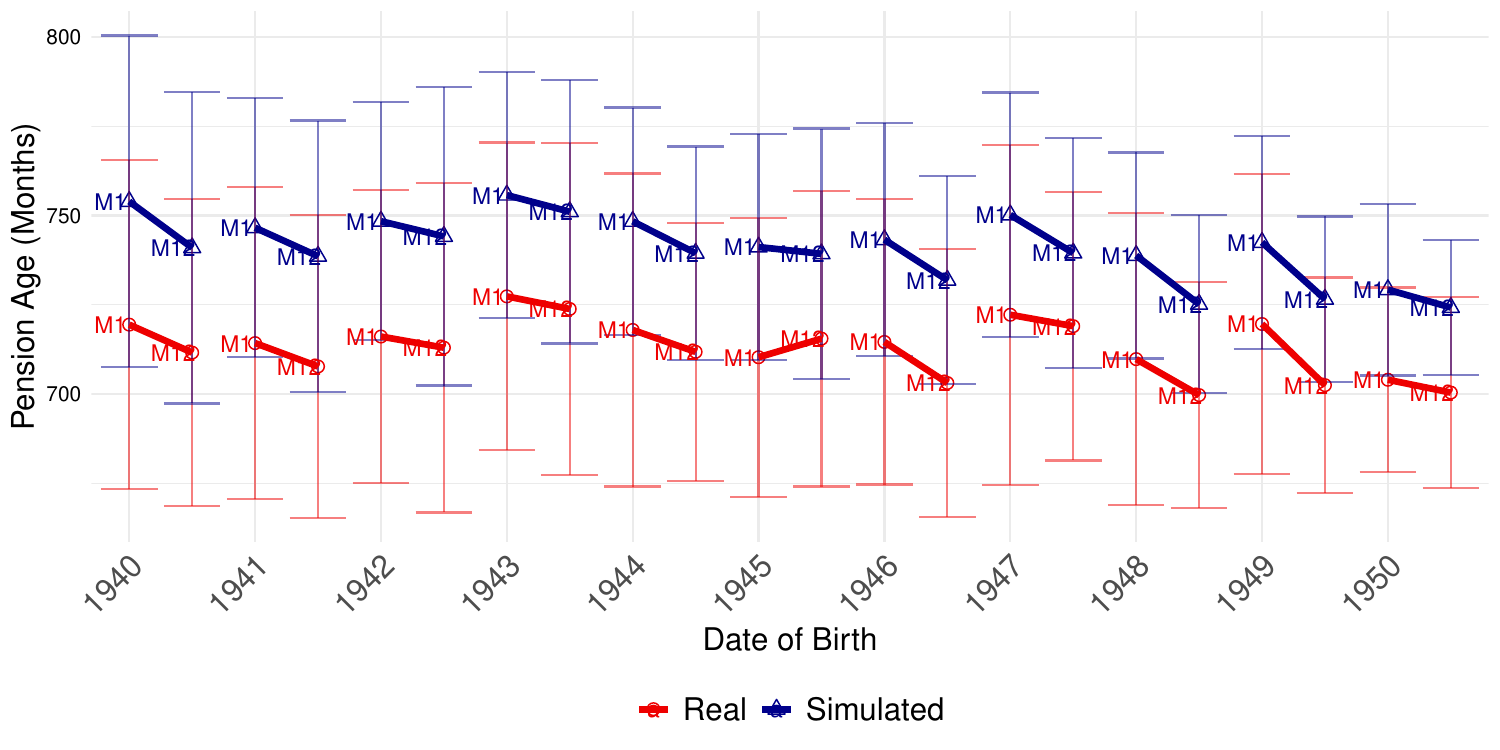}
\caption{Real (circle dots) and model-generated (triangle dots) average retirement ages, measured in months, for male individuals born between 1940 and 1950, using a cutoff one year before the observed retirement year. Each point indicates the cohort-specific average retirement age for January (M1) and December (M12), with vertical bars representing one standard deviation.}
\label{fig:pension_result_decoder_only}
\end{figure}

Figure \ref{fig:pension_result_decoder_only} shows results with the one-year cutoff. Here, the model reproduces the negative effect of December birth on retirement age, indicating robustness of the signal. However, a clear upward bias emerges. This suggests that conditioning the model too close to the event may shift its internal expectations and effectively push retirement further into the future.  

\begin{figure}[!htbp]
\centering
\includegraphics[width=\columnwidth]{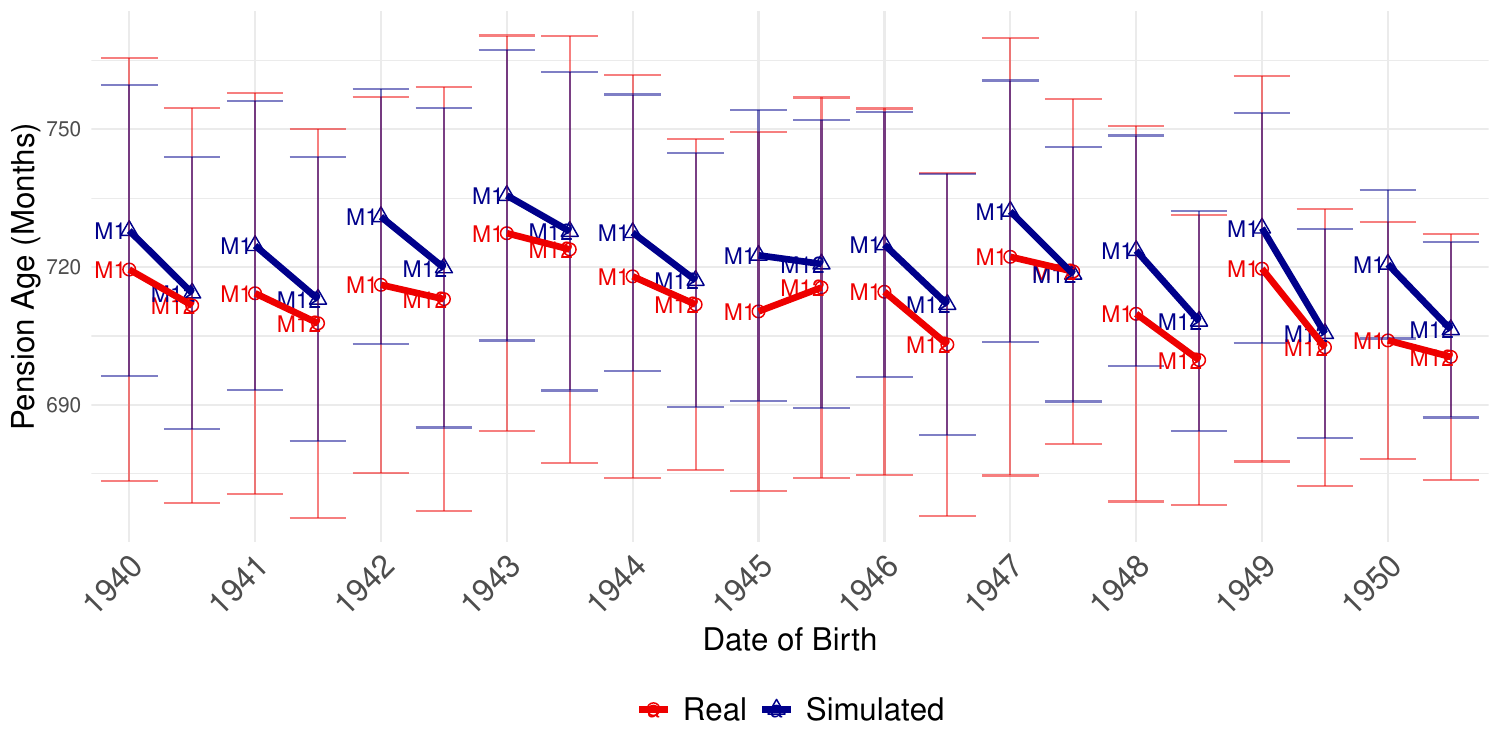}%
\caption{Real (circle dots) and model-generated (triangle dots) average retirement ages, measured in months, for male individuals born between 1940 and 1950, using a cutoff four years before the observed retirement year. Each point indicates the cohort-specific average retirement age for January (M1) and December (M12), with vertical bars representing one standard deviation.}
\label{fig:pension_result}
\end{figure}

Figure \ref{fig:pension_result} reports results with the four-year cutoff. The model captures the expected empirical pattern: December-born individuals retire earlier than January-born peers, and the magnitude of this birth-month effect is well preserved across cohorts. The upward bias in retirement ages is largely absent under this earlier cutoff, indicating that conditioning further from the retirement event leads to a closer alignment with the observed timing of pension entry.

Table~\ref{ate} formalises these findings. Both cutoffs generate significant differences in retirement age between treated (born in January) and untreated (born in December) individuals, consistent with the empirical benchmark. Yet the size of the model's effect is sensitive to the conditioning horizon. 
Truncating the sequence further from the outcome offers a stronger test of counterfactual reasoning. Still, it tends to overstate effect sizes, illustrating how the length of the conditioning horizon shapes both predictive accuracy and the model’s interpretation of the causal nexus among life events.



\subsection{Out-of-sample Generation} \label{out_of_sample}

We evaluate the model’s out-of-sample performance by investigating the coherence of generated life sequences beyond the temporal window of the training data, i.e., after 2015. In this setting, the model is conditioned on observed sequences up to 2015 and generates tokens into the future. Good out-of-sample performance indicates robustness, a desirable property for a model expected to simulate plausible scenarios.


Given the absence of ground truth data for these later years, our evaluation focuses on syntactic and structural correctness rather than predictive accuracy. Specifically, we examine the generated sequences for internal consistency, valid token ordering, and the absence of structural violations, such as overlapping events or misaligned episode boundaries (more details in Appendix~\ref{oos_gen}). This assessment allows us to verify that the model preserves the learned ``grammar'' of life sequences when extrapolating beyond the observed data.

Figure~\ref{fig:out_of_sample_decoder_only} presents the density of out-of-sample generated years in which the first structural failure occurs, considering only sequences where a failure did occur within a 20-year generation window. The results show that sequence generation begins to fail around year 7 or 8, though a non-negligible share persists further, with successes becoming exponentially less likely as the generation horizon extends, and 5\% of sequences remain structurally valid after 20 years.

\section{DISCUSSION} \label{discussion}

We propose a novel encoding method for converting longitudinal administrative data into sequences and train a generative life sequence Transformer model. Despite its relatively small size, the model learns to reproduce realistic and causally consistent labour market histories and generates coherent synthetic life paths conditional on past trajectories, showing the potential of Transformer architectures in this domain.

Through empirical exercises, we study the strengths and limitations of our model. On the one hand, it successfully reproduces causal patterns, from birth-month effects on retirement to the motherhood penalty. On the other hand, the exercises reveal systematic sensitivities. Rule-driven discontinuities are smoothed over when the underlying events are rare in the data; treatment effects vary depending on how counterfactuals are defined, whether by comparing mothers to non-mothers or by truncating retirement histories at different horizons; and the timing of effects can shift, with the model sometimes understating short-run penalties but prolonging them over the longer term. 

These findings highlight a broader challenge: counterfactual inference with generative models depends on how we define treatment, select controls, and decide where to cut the input history. These are not just technical choices but conceptual ones, raising fundamental questions about what constitutes a life sequence counterfactual. They also suggest future research on developing principled methods for introducing researcher-imposed variation into simulated life sequences, ensuring theory-guided design of counterfactual scenarios. 

Finally, while our results show encouraging alignment with empirical findings, they also remind us that predictive modelling of individual life courses carries ethical and societal risks. Structural biases present in training data will be reinforced. Without appropriate human oversight, there is also the potential for misuse, for example, in automated decision-making systems that influence resource allocation or eligibility for benefits. Future applications must combine methodological advances with safeguards for fairness, privacy, and responsible use. 

\subsubsection*{Acknowledgements}

The authors acknowledge European NextGenerationEU funding through the PNRR project ``Age-It - Ageing well in an ageing society'', code PE0000015, MUR 341/2022 tender.

\bibliography{references}
\bibliographystyle{abbrvnat}

\clearpage
\appendix
\thispagestyle{empty}

\onecolumn
\aistatstitle{Supplementary Materials}

Counterfactual inference is a cornerstone of scientific reasoning in disciplines such as political science \cite{sekhon2009opiates}, 
economics \citep{heckman2008econometric}, history \citep{levy2015counterfactuals},  and psychology \citep{foster2010causal}, and has also sparked ongoing methodological debate \citep{cartwright200710}. Counterfactual identification requires constructing hypothetical scenarios that resemble what would have happened in the absence of an event assumed to be causal \citep{Rubin01032005}. Such a scenario allows quantifying the impacts of specific events, like treatments or policies, or the underlying causes of observed outcomes. Typical research questions include: What is the causal effect of increasing the minimum wage on household consumption \citep{alonso2022beyond}? What are the drivers of civil conflict \citep{acemoglu2020population}?

The gold standard approach for identifying causal effects involves randomised interventions, in which the treatment is assigned randomly to some individuals and withheld from others \citep{cartwright2007rcts}. The difference in outcomes between the treated and control groups is interpreted as the causal effect of treatment. Randomisation ensures that characteristics that might correlate with the outcome of interest, such as gender, age, and socioeconomic status, are, on average, evenly distributed across the treatment and control groups. Thus, any systematic outcome difference can be attributed to the treatment itself.

However, randomisation rarely occurs naturally and can be challenging to implement even in controlled settings, due to ethical constraints and substantial costs \citep{thomas2016after}. As a result, economists and other researchers have developed strategies and mathematical models to identify suitable control groups without relying on random treatment assignment. Despite being widely adopted, these observational methods often rely on specific assumptions about the data and the underlying causal mechanisms, which are typically difficult to verify in practice. 

\section{Difference-in-differences} \label{dif_in_dif}

Difference-in-differences (DiD) is one of the most commonly used causal identification strategies for evaluating the effects of policy interventions by comparing changes in an outcome variable between a treatment group and a control group, both before and after the intervention. Here, the main assumption is that the treated and control groups would have followed parallel trends over time in the absence of the policy or event under evaluation. This assumption, while critical, is rarely validated and often debated in empirical applications.  \par

\cite{card2000minimum} on minimum wage in New Jersey is considered one of the most famous DiD studies; the authors compared employment in the fast food sector in New Jersey and in Pennsylvania, in February 1992 and in November 1992, after New Jersey's minimum wage rose from \$4.25 to \$5.05 in April 1992. Observing a change in employment in New Jersey only, before and after the treatment, would fail to control for omitted variables such as weather and macroeconomic conditions of the region. By including Pennsylvania as a control in a difference-in-differences model, any bias caused by shared variables between New Jersey and Pennsylvania is implicitly controlled for, even when these variables are unobserved. Assuming that New Jersey and Pennsylvania have parallel trends over time, Pennsylvania's change in employment can be interpreted as the change New Jersey would have experienced had it not increased the minimum wage, and vice versa. The evidence suggested that the increased minimum wage did not lead to a decrease in employment in New Jersey, contrary to what some economic theories would suggest.

\subsection{Event Study} \label{event_study}

While powerful, DiD is not always applicable, especially in settings where no natural control group exists. Consider, for example, the question: What is the causal effect of childbirth on a woman’s career trajectory? Ideally, one would randomise fertility to estimate this effect, but such an experiment is infeasible. Applying DiD would require identifying a group of women identical to mothers in both observable and unobservable characteristics, and then assuming their career paths would have evolved similarly, absent childbirth. In practice, achieving such a match is highly challenging.

In such cases, researchers often turn to an event study design \citep{jacobson1993earnings}, a generalisation of DiD that exploits variation in treatment timing. Event studies are especially useful when all units are eventually treated at different times. The method compares changes in outcomes before and after treatment for each unit, leveraging variation across units treated at different points in time. In the childbirth example, this allows for estimating treatment effects by aligning individuals around the “event” (birth of a child) and comparing pre- and post-treatment trajectories. Like DiD, event studies rely on a version of the parallel trends assumption—that, absent the event/treatment, the trend of the outcome would have continued as observed in the pre-treatment period.

\subsection{Propensity Score} \label{propensity_score}

While Difference-in-Differences (DiD) relies on the parallel trends assumption, in many empirical settings, treatment and control groups differ substantially in observable characteristics, making the assumption less credible. One strategy to mitigate this concern is to combine DiD with Propensity Score Matching (PSM), a method introduced by \cite{rosenbaum1983central}. The core idea of PSM is to construct a control group that is statistically similar to the treated group in terms of observed covariates, by matching treated and untreated units with comparable estimated probabilities of treatment (the “propensity score”).

When PSM is combined with DiD, the matching procedure reduces bias from differences in observable characteristics, while the DiD framework accounts for unobserved time-invariant heterogeneity. This hybrid approach is particularly valuable when treatment assignment is not random and when untreated units differ systematically from treated ones. By matching first and then applying DiD, researchers ensure that the parallel trends assumption is more plausible for the reweighted or matched sample.

\subsection{Synthetic Control} 

The Synthetic Control Method (SCM) is another causal inference approach closely related to difference-in-differences. It is particularly well-suited to cases where a single unit (such as a country, region, or firm) undergoes treatment, and no comparable control group exists. Instead of relying on a parallel trends assumption between predefined groups, SCM constructs a synthetic version of the treated unit using a weighted combination of untreated units. These weights are chosen to closely match the treated unit’s outcome trajectory and covariates in the pre-treatment period.

This method was originally proposed by \citet{abadie2010synthetic} to study the economic impact of California’s 1988 tobacco control program. Rather than comparing California to a single state, SCM used a weighted average of other U.S. states to construct a “synthetic California” that closely mirrored its economic indicators before the intervention. Any divergence between the treated unit and its synthetic control after treatment can then be attributed to the intervention, under the assumption that the synthetic control approximates the counterfactual trajectory the treated unit would have followed in the absence of treatment.

Compared to standard DiD, SCM is more flexible and data-driven. It relaxes the requirement that the treatment and control units follow identical trends and instead ensures a close match in the pre-treatment period. This makes SCM particularly valuable in comparative case studies or when few treatment units are available, though it also comes with limitations—such as sensitivity to model specification and the requirement for a rich pool of potential control units.

\section{Work History Italian Panel Data} \label{whip_data}



In Italy, the organisation that collects social insurance contributions and distributes social security benefits, such as pensions, maternity leave, and unemployment benefits, is the Istituto Nazionale della Previdenza Sociale  (INPS). As such, INPS gathers data on significant workforce segments, both active and inactive. The reference population comprises all individuals employed by private firms, workers with quasi-dependent employment arrangements, self-employed, retirement spells, and non-working spells in which the individual received social benefits, like unemployment subsidies or mobility benefits. The workers for whom activity is not observed are those who worked as freelancers and have an autonomous security fund. The observed group represents all manufacturing, construction, and services production sectors. However, it does not cover public employment, which involves approximately 3.5 million people, predominantly in the education and health sectors. Certain sectors, such as agriculture, are not fully represented in INPS data since its data does not cover any activity in the shadow economy. If an individual is observed working one year in a private company, and then he/she start working in the shadow economy, INPS will register only that one year the individual was legally working. 

The Work History Italian Panel (WHIP) \citep{leombruni2010note, bena2012new} consists of a systematic sample of 1 in 15 individuals drawn from the INPS database based on their date of birth. This database allows us to track the employment histories of sampled individuals, including all periods of employment recorded by INPS, retirement, and any time spent receiving social security benefits, such as unemployment subsidies or maternity leave. The WHIP currently covers the period from 1985 to 2019.

For our analysis, we limit the sample to WHIP observations recorded from 1990 to 2015 for training. 
We started in 1990 as the information's reliability improved significantly from that date onward. Additionally, we apply the following selection criteria to further refine our sample:

\begin{enumerate}

\item Have information until the end of the observed period (2015) 
\item Effective period observed greater than or equal to half of the worker's potential period. Thus, we include only individuals observed for at least half of their maximum possible participation period, where labour force participation is assumed to start at age 30 and is bounded by the 1990–2015 data window and an 85-year lifespan. For example, an individual born in 1975 turns 30 in 2005, we expect to observe him/her from 2005 to 2015. If we observe him only from 2013 to 2015 (3 years), the ratio between effective (3) and potential (10) is 0.3. Therefore, we exclude this individual from the sample. 

\item No missing date of birth
\item At least five records
\item Age of entry in the panel higher than 15 and lower than 80. We exclude those individuals who are too young or too old.
\item Fraction of returns or investment income over the total number of records lower than 90\%. We exclude people who are fully detached from the workforce.

\begin{table}[h]
\centering
\begin{tabular}{|c|c|}
\hline
   Initial Number of Individuals & 4032674 \\ \hline 
  \textbf{Selection Criteria}   & \textbf{\% Drop} \\\hline
  1. & 13.24\\
  2. & 16.45\\
  3. & 0.16\\
  4. &  6.5\\
  5. & 4.91\\
  6. & 13.83\\ \hline
Final Number of Individuals & 1652877 \\ \hline  
\end{tabular}
\caption{Sample selection. Each row in the table corresponds to a selection criterion. The second column reports the percentage of individuals discarded, regardless of the other selection criteria.}
\end{table}
\end{enumerate}

\subsection{Labour Records} \label{whip_labour}

The primary individual information in WHIP is yearly income. This is the most important information for INPS to compute contributions and monitor social security benefits. Many other variables are attached to each yearly income. For our study, we focus on 15 income-related variables.

\paragraph{Demographic information} WHIP records include an individual's demographic/background information. These are (1) sex, (2) birth date, and (3) macro-area of birth. In our case, sex is a binary variable that encodes male or female. We retrieve only the birth year from the birth date to encode the age when a labour event occurs. The macro-area of birth specifies one of the 5 Italian groups of regions (North-East, North-West, Center, South, or Island). There is a sixth category for foreign-born people.

\paragraph{Workplace information} Another set of variables pertains to features of the workplace. Notably, only for working individuals the following information is stored, e.g., residents receiving a pension, unemployment benefits or rents, do not have any of these specified, while for a self-employed person we observe only the workplace location.  

(4) Economic activity/Sector denotes the services or goods that the company produces. WHIP encodes this information through the ATECO classification system\footnote{\url{https://codiceateco.it/sezioni}}. ATECO is the national version of the European nomenclature, Nace Rev. 2, Regulation (EC) No 1893/2006 of the European Parliament and of the Council of 20 December 2006\footnote{\url{https://eur-lex.europa.eu/eli/reg/2006/1893/oj/eng}}. The ATECO has a hierarchical tree-like system, each additional digit provides a more detailed breakdown of the service, see the example in the following table. For our study, we use only the first four digits of the ATECO code to balance informativeness and sample size of each category. 

\begin{table}[h]
\centering
\begin{tabular}{|c|c|}
\hline
  \textbf{ATECO CODE}   & \textbf{Definition} \\
  \hline
   G  & Wholesale and retail trade\\
   45 & Repair of motor vehicles and motorcycles\\
   45.1 &Trade in motor vehicles\\
   45.11 & Trade in passenger cars and light motor vehicles\\
\hline
\end{tabular}
\end{table}

(5) Enterprise size is encoded in 5 ordinal categories: less than 10 employees, between 10 and 19, between 20 and 199, between 200 and 999, and more or equal to 1000. Finally, (6) Workplace location can be one of the 107  Italian provinces, plus an additional category for foreign locations. 

\paragraph{Individual income information} The last set of features provides details about the individual's income and its nature. (7) Labour force status is specific to WHIP, i.e., it does not follow any international classification system. It describes the type of income, i.e., in what context the income is associated with a person. Examples of these statuses include: Employee, Pensioner, Self-employed, Unemployed, encompassing work-related and other socioeconomic statuses. Employee, Self-Employed, or Para-Subordinate (e.g., sub-contractors, consultants, people with work contracts but not all the benefits) have one additional modifying information, the (8) Working title which indicates the qualification in the job and may fall into one of 11 categories: in training, Workman, Clerk, Junior Manager, Senior Manager, technical-administrative, Sportsman, Artist, Voucher, Collaborator, or Professional. \par 

For Employee records only, we have four additional modifiers: (9) work intensity, (10) maternity leave intensity, (11) sick leave intensity, and whether the work is (12) part-time or full-time. 
The three intensities describe the average number of weeks in a month when an employee was working, on maternity leave, or sick. We derive the intensity by dividing the overall number of weeks spent in a given category (e.g., working) by the total number of months the employment relationship lasted in a given year. Therefore, each record can take a maximum value of around 4.3 (52 weeks divided by twelve), and a minimum value of 0. Notably, maternity leave is only recorded for female employees. Finally, we discretise the intensity into five categories based on the rounded value. If, for example, the intensity work is zero, we encode it as \texttt{<WRKINT\_S0>}, if the intensity is less than 1, we encode it as \texttt{<WRKINT\_S1>}, for intensities greater than 4, they are encoded into the category \texttt{<WRKINT\_S4+>}.

The final attributes are the (13) month the event started, its (14) duration in months within a year, and (15) income during the event. The nature of income differs depending on the labour force status. For example, for Employees, Self-employed, Para-Subordinate, and Rentiers, it is the reported income from which the INPS calculates social security contributions. For Unemployed and Pensioner, the income is the amount that INPS provides to the individual. The raw version of the income is a yearly value; we divide it by the duration in months of a given labour force status within a year and derive a monthly income. We further adjust it for inflation, deriving a real monthly income, and discretise this value into 100 quantiles. 




\begin{table}[h!]
\centering
\begin{tabular}{|c|c|c|}
\hline
  \multirow{4}{*}{General} &  Number of Individuals & 1652877  \\
  & Average number of events & 30.48 \\
  & Median Year &  2005 \\
  & Median Age & 42 \\\hline
  \multirow{8}{*}{Background} &  \% Female &  37.1\\
    & \% North-West  & 22.6\\
    & \% North-East & 18.4\\
    & \% Center & 15.9\\
    & \% South  & 22.4\\
    & \% Islands& 9.73\\
    & \% Foreign born  & 11\\
    & \% Missing Area of Birth& 0.0005\\
   & Median Year of Birth  & 1966 \\\hline
\end{tabular}
\caption{Descriptive Statistics of the Dataset}
\end{table}

\begin{figure}[!thp]
\centering
\includegraphics[width=1\columnwidth]{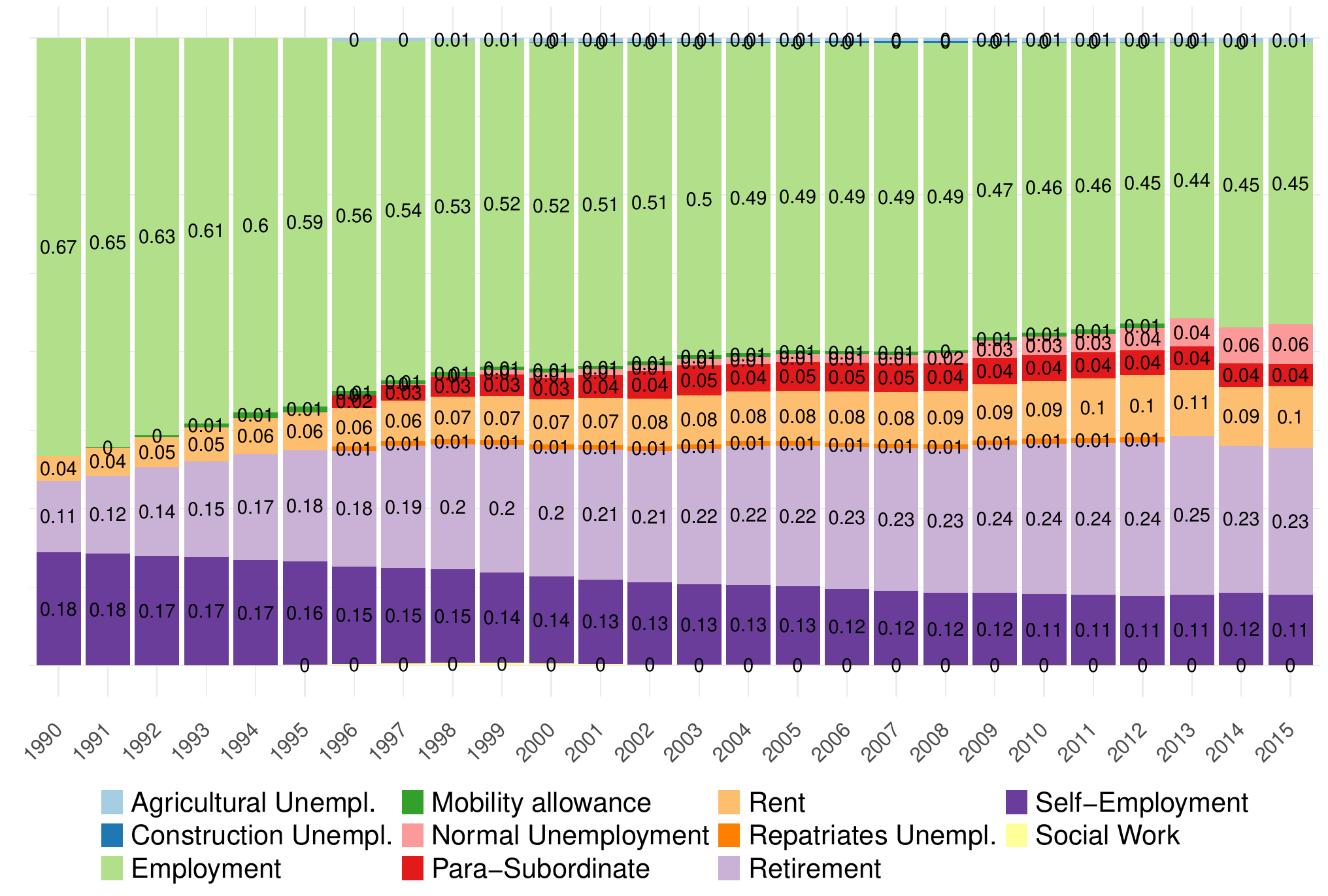}%
\caption{Labour Force Status composition for each year in the global dataset}
\floatfoot{}
\label{fig:labour_force_status_composition}
\end{figure}

\paragraph{Vocabulary}

The collection of all concept-tokens creates the vocabulary $\mathcal{V}$. The vocabulary is made up of tokens for all categories of each variable and has size $\mid \mathcal{V} \mid $ = 661. The vocabulary  $\mathcal{V}$ is generated from the training subset and does not include all the possible categories in the data. Table \ref{vocabulary} lists all variables and their possible categories.

\begin{table}[h!]
\centering
\begin{tabular}{|c|c|c|c|}
\hline
  \textbf{Token Type}   & \textbf{Variables} & \textbf{N. Categories}  & \textbf{Encoding} \\
  \hline
  \multirow{3}{*}{Background Information} & Year of Birth & 86  & 1913-1998\\
  & Sex & 2 & F-M\\
  & Area of Birth & 6  & 1-6 (Foreign)\\\hline
 \multirow{12}{*}{Labour} & Enterprise type & 280  & ATECO 4-digits\\
 & Enterprise size & 5  & WHIP classification\\
 & Work Location &   110 & Italian provinces codes\\
 & Labour force status &   11 & WHIP classification\\
 & Work title &   11 & WHIP classification\\
 & Work Arrangements &   2 & Part time - Full time\\
 & Work intensity &   5  &  Threshold based \\
 & Maternity leave intensity &   5  & Threshold based\\
 & Sick leave intensity &   5  &Threshold based \\
 & Income &   100  & Quantile based \\
 & Start Month & 12& 1-12\\
 & Duration & 12& 1-12\\\hline
\multirow{1}{*}{Special} & Special &11& [PAD]...[UNK] \\
\hline

\end{tabular}
\caption{Vocabulary specification}
\label{vocabulary}
\end{table}

\paragraph{Labour life sequence}

Each life sequence is represented as a list of discrete tokens, encoding time-indexed events and their characteristics. Sequences are capped at a maximum of 1,560 tokens. Any sequence shorter is padded with a padding token, and any sequence longer is truncated starting from the earliest years (i.e., from the start of the sequence), ensuring that all input tensors are of consistent shape for training and inference. Longer sequences are truncated by whole years so as not to lose the temporal structure of the sequence (i.e., all tokens in a year are deleted until the sequence is at most 1,560 tokens long). 

When an individual has no recorded activity in a given year, the year is still represented in the sequence. We then insert a silent event placeholder: \texttt{<MONTH\_1>} followed by \texttt{<DURATION\_12>} with no attributes in between, to preserve the temporal regularity of the sequence.
The ordering of tokens within a sequence is chronological. Events that begin in the same calendar month are randomly shuffled relative to each other, regardless of their duration. This is based on the assumption that events co-occurring in the same month cannot be reliably ordered based on observed data, and shuffling them prevents the model from overfitting to arbitrary patterns in event ordering. Outside of these same-month shuffles, the rest of the sequence remains ordered.
Besides same-month shuffling, we introduce a small amount of token dropout, randomly removing about 1\% of tokens during training. This regularisation mechanism discourages the model from overfitting to missing data and promotes robustness. The dropout is uniformly applied and does not target any specific token type.

Inside each event, there is a meaningful and deliberate order to the tokens. Their order reflects a semantic priority derived from the downstream modelling goals. For example, income is placed before ATECO (economic activity), because for many of our tasks, we are more interested in how income can inform or predict the sector of employment, rather than the reverse. That is, we assume income to be more causally upstream in our use cases. While alternative orderings are possible, changing the token order within events would require retraining, as the model implicitly learns positional correlations. Nevertheless, this ordering can be adjusted in future work depending on the application.

The sequence lengths (measured as the number of tokens per individual life trajectory, excluding background information) have the following descriptive statistics

\begin{table}[h]
\centering
\begin{tabular}{|l|ccccc|}
\hline
  & \textbf{Mean}   & \textbf{Median} & \textbf{Mode}& \textbf{Minimum } & \textbf{Maximum} \\
  \hline
   Tokens & 232  & 220 &  128  & 21 &1554 \\
\hline
\end{tabular}
\caption{Descriptive statistics of sequences}
\end{table}




\textbf{This example is not based on an individual in the dataset, any resemblance to a real individual is coincidental.} Consider an example as an illustration of the encoding:

\texttt{A3  | F | MONTH\_1  | YEAR\_1942}, 

i.e., a female born in January of 1942 in the Center region of Italy. Assume that in the first year of measurement (1990), she worked full time during the
whole year, then had a year of no recorded activity, and then got another job in a different province for the entire following year:

\texttt{
[BOL] (0, 0) | \textcolor{brown}{MONTH\_1} (1, 48) | \textcolor{blue}{[TIPO\_1]} (1, 48) | INCOME\_20 (1, 48) | WRKT\_1 (1, 48)  | WRKP\_XX (1, 48)  | ATE\_A123 (1, 48)  | FSIZE\_1 (1, 48)  | FULL\_TIME (1, 48)  | WRKINT\_S3 (1, 48)  | SIKINT\_S0 (1, 48)  | MATINT\_S0 (1, 48)  | \textcolor{brown}{DUR\_12} (1, 48) | \textcolor{orange}{[EOY]} (1, 48) |  \textcolor{brown}{MONTH\_1} (2, 49) | \textcolor{brown}{DUR\_12} (2, 49) | \textcolor{orange}{[EOY]} (2, 49)|
\textcolor{brown}{MONTH\_1} (3, 50) | \textcolor{blue}{[TIPO\_1]} (3, 50) | INCOME\_22 (3, 50)  | WRKT\_2 (3, 50)  | WRKP\_YY (3, 50)  | ATE\_A124 (3, 50)  | FSIZE\_1 (3, 50)  | FULL\_TIME (3, 50)  | WRKINT\_S3 (3, 50)  | SIKINT\_S0 (3, 50)  | MATINT\_S0 (3, 50)  | \textcolor{brown}{DUR\_12} (3, 50) | \textcolor{orange}{[EOY]} (3, 50)
}

The numbers inside parentheses next to the tokens are the year (measured as years after 1989) and her age. Notice that the \texttt{[EOY]} token is measured with the current year and age, and once this token appears, the year and age increase by one. Hence, when the model generates an \texttt{[EOY]} token, time embeddings are updated deterministically. Each of her working spells is different because of its attributes, though for both of her working spells, she worked an average of 2 to 3 weeks each month and never took sick or maternity leave. 

Her life sequence continues for every year until 2015, and when she is around 60 years of age, she retires and starts getting a pension from INPS:

\texttt{
\textcolor{orange}{[EOY]} (12, 59) | \textcolor{brown}{MONTH\_1} (13, 60) | \textcolor{blue}{[TIPO\_1]} (13, 60) | INCOME\_39 (13, 60)  | WRKT\_2 (13, 60)  | WRKP\_YY (13, 60)  | ATE\_A124 (13, 60)  | FSIZE\_1 (13, 60)  | FULL\_TIME (13, 60)  | WRKINT\_S3 (13, 60)  | SIKINT\_S0 (13, 60)  | MATINT\_S0 (13, 60)  | \textcolor{brown}{DUR\_5} (13, 60) | \textcolor{brown}{MONTH\_5} (13, 60)  | \textcolor{blue}{[TIPO\_10]} (13, 60)  | INCOME\_35 (13, 60)  | \textcolor{brown}{DUR\_8} (13, 60) | \textcolor{orange}{[EOY]} (13, 60)  | \textcolor{brown}{MONTH\_1} (14, 61)  | \textcolor{blue}{[TIPO\_10]} (14, 61)  | INCOME\_35 (14, 61)  | \textcolor{brown}{DUR\_12} (14, 61) | \textcolor{orange}{[EOY]} (14, 61) 
}

Notice she still works for the first 5 months of 2005 and only starts receiving her pension in May. After that, she is likely to stay on a pension scheme until 2015. 

\section{Model architecture}

Transformers are a class of neural network architectures initially developed for natural language processing tasks, such as machine translation or text summarization \citep{vaswani2017attention}, but are now being applied to a wide variety of sequence types from different domains, including protein discovery \cite{consens2025transformers}, health predictions \citep{yang2023transformehr}, music generation \cite{agostinelli2023musiclm}. Their strength lies in the self-attention mechanism, which enables them to model long-range dependencies in sequential input. 

One of the most influential \textit{encoder-only} Transformer models is BERT (\textit{Bidirectional Encoder Representations from Transformers}), introduced by \citet{devlin2019bert}. BERT is trained using a masked language modelling (MLM) objective, where a subset of tokens in the input sequence is randomly masked, and the model learns to predict them using bidirectional context, i.e., both left and right tokens. This bidirectional attention mechanism enables BERT to build rich, context-aware representations of entire sequences, making it highly effective for discriminative tasks such as text classification, question answering, and semantic similarity. Since BERT processes the entire input at once, it excels at tasks that require a holistic understanding of the input sequence, rather than step-by-step prediction. An \textit{encoder-only} version of our model is similar to \cite{savcisens2024using}'s Life2Vec, which is inspired by the BERT architecture and training objective.

\begin{figure}[!htbp]
\centering
\includegraphics[width=\columnwidth]{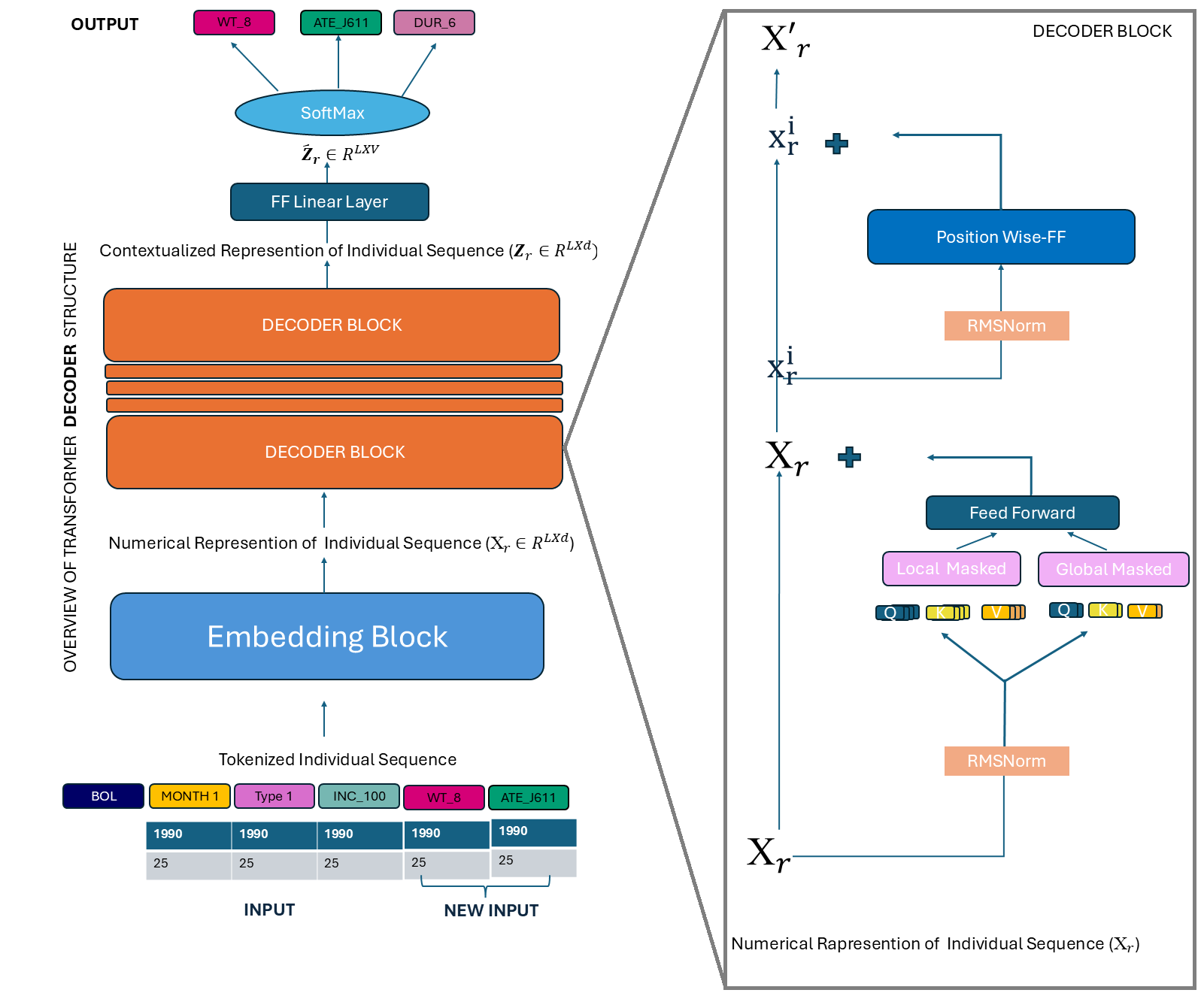}
\caption{Overall architecture of our decoder-only model. Each tokenised life sequence is passed through an embedding block and a stack of decoder blocks. Each token, consisting of concept, age, and year information, is embedded using three components: concept embeddings retrieved from a learned lookup matrix $\mathcal{E}_\mathcal{V}$ (normalised by subtracting its mean), and temporal features (age and year) encoded via Time2Vec. These are combined with learnable positional embeddings to form the input token representations. The resulting sequence is processed through multiple decoder blocks, each structured to preserve the autoregressive nature of the task. Each block starts with RMS normalisation, followed by a masked multi-head attention layer (using the Performer mechanism with local and global heads), ensuring that each token only attends to preceding tokens. The output of the attention layer is combined with the input via a residual connection. This is followed by another RMS normalisation and a position-wise feed-forward layer, whose output is again added through a residual connection. Finally, the contextualised output representations are passed through a linear projection head and to a softmax function that assigns a probability to each token in the vocabulary. The predicted token is the one with the highest probability score.}
\label{fig: decoder only}

\end{figure}

In contrast, GPT (\textit{Generative Pre-trained Transformer}) exemplifies a \textit{decoder-only} architecture. Introduced by \citet{radford2018improving}, GPT is trained with an auto-regressive objective, predicting the next token in a sequence based solely on the preceding tokens. GPT uses causal (masked) self-attention, which restricts each token's attention to preceding positions in the sequence, preventing information from ``leaking'' from future tokens. Unlike BERT, GPT does not explicitly mask input tokens, but the causal attention mechanism ensures that generation proceeds in a left-to-right manner. This makes the GPT architecture particularly well-suited for generative tasks, such as language generation, sequential modelling, and, in our case, simulating counterfactual life sequences. Figure \ref{fig: decoder only} illustrates the decoder-only architecture we train in this paper. 

\newpage

\section{Experimental details}\label{experimental_details}

This section outlines the experimental setup, notation, and estimands, as well as the process of converting generated simulations into treatment effects that are directly comparable to their empirical counterparts. Our goal is to place generative model-based evaluation on the same footing as standard causal inference designs in applied economics. We build on the event study and propensity-score frameworks described in Appendix \ref{event_study}--\ref{propensity_score}, adapting them to the context of sequences generated from our model.

\paragraph{Common setup and notation}

Let $i$ index selected individuals in the test set. Each empirical exercise is framed in terms of a binary treatment $T_i$ and a scalar outcome $Y_i$. The central estimand is the Average Treatment Effect (ATE), defined as the expected difference in outcomes between treated and untreated/control units:
\[
\text{ATE} = \mathbb{E}[Y_i(1) - Y_i(0)],
\]
where $Y_i(1)$ and $Y_i(0)$ denote the potential outcomes for individual $i$ under treatment and control, respectively. In practice, we never observe both potential outcomes for the same person; however, under appropriate research designs (such as random assignment, natural experiments, or valid identification strategies), we can estimate the ATE as the difference in mean outcomes between the treated and control groups. In our experiment, these quantities are:

\begin{itemize}
    \item Pension: $T_i = 1$ if individual $i$ was born in January, $T_i = 0$ if born in December. The outcome $Y_i$ is the age at first pension claim, measured in months.
    \item Unemployment: $T_i = 1$ if individual $i$ experienced their first unemployment spell at or after 40 years of age (480 months), $T_i = 0$ otherwise. The outcome $Y_i$ is the number of months spent in the first unemployment spell. The estimand is local, defined within a symmetric window $h$ around the 480-month cutoff.
    \item Maternity: $T_i = 1$ if individual $i$ experiences at least one maternity leave episode in her sequence, $T_i = 0$ otherwise. The outcome $Y_i$ is the average post-maternity annual earnings.
\end{itemize}

For maternity, treatment is endogenous rather than rule-based. As such, the definition of a control group is non-trivial. Thus, we employ an event study design and a propensity score matching approach (Section \ref{event_study}) to align careers around childbirth events and identify a suitable control group of non-mothers that matched he pre-birth characteristics of the mothers. To make results comparable across experiments, we summarise the estimated time-path into a single number: the mean effect over the birth year and the first ten post-birth years (see Section \ref{mothers}).

\paragraph{From simulations to outcomes}

\begin{figure}[h]
  \begin{subfigure}[t]{0.5\textwidth}
  \centering
\includegraphics[scale = 0.4]{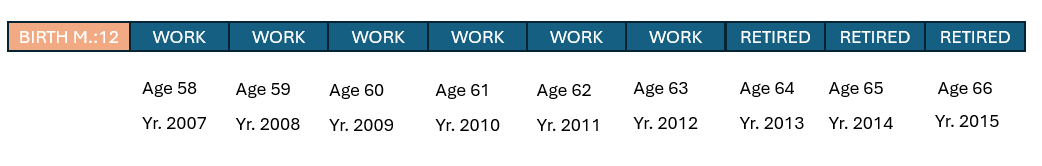}
\caption{Original sequence}
\end{subfigure}
\vfill
\begin{subfigure}[t]{0.4\textwidth}
\includegraphics[scale = 0.4]{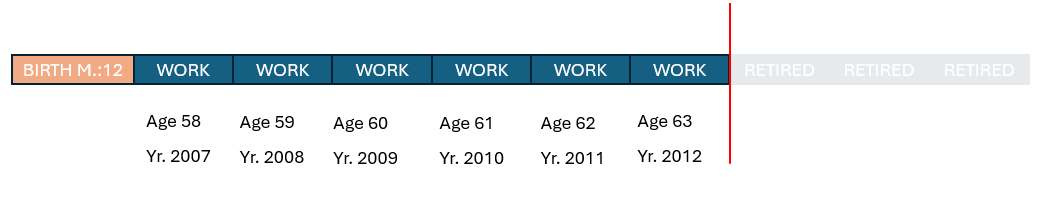}
\caption{Sequence Cut}
  \end{subfigure}
      \hfill
\begin{subfigure}[t]{0.4\textwidth}
\includegraphics[scale = 0.4]{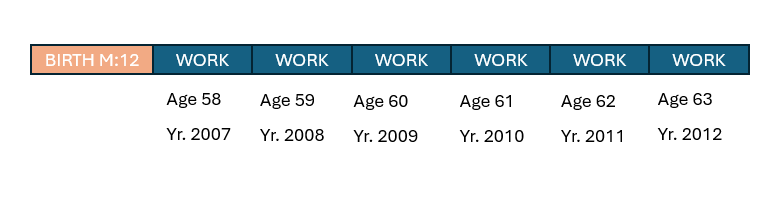}
\caption{Final input sequence}
 \end{subfigure}
 \caption{Example of the simulation setup: (a) shows the original work sequence, (b) the sequence cut applied one year before retirement occurs, and (c) the resulting input sequence used for simulation.}
 \label{fig:sequence_cut}
\end{figure}

For every person $i$ we generate $K$ continuations from a fixed cutoff of their observed history. This cutoff point corresponds to the moment just before the outcome of interest takes place:  

\begin{itemize}
    \item Pension: one year before observed retirement (denoted 1y in tables) or four years before retirement (4y). An example of the former setup is shown in Figure \ref{fig:sequence_cut}.
    \item Unemployment: the year of the first observed unemployment spell.  
    \item Maternity: the year of the first observed maternity leave (offset--0), as well as one year prior (offset--1), to allow counterfactual generation without maternity.  
\end{itemize}

Each simulation $k$ yields a synthetic outcome $\tilde Y_{ik}$. We compute the person-level Monte Carlo mean

\[
 Y_i^{\text{sim}} \;=\; \frac{1}{K}\sum_{k=1}^K \tilde Y_{ik},
\]

and then use $\{Y_i^{\text{sim}}\}_{i=1}^n$ to estimate treatment effects in the same way we do with real outcomes $\{Y_i^{\text{real}}\}_{i=1}^n$. This ensures that all ATEs are defined at the individual level, not at the simulation-draw level. 



We report sample sizes $n$, number of simulations $K$, and uncertainty measures obtained via a paired bootstrap at the individual level. Specifically, we resample individuals (with replacement) from the fixed analysis sample, e.g., January or December retirees for pensions, or individuals within the 40-year bandwidth for unemployment. For every bootstrap sample, we recompute both the empirical and model ATEs,
\[
\widehat{\text{ATE}}_{\text{emp}}=\mathbb{E}[Y_i^{\text{real}} \mid T_i{=}1] \;-\; \mathbb{E}[Y_i^{\text{real}} \mid T_i{=}0], \qquad 
\widehat{\text{ATE}}_{\text{model}}=\mathbb{E}[Y_i^{\text{sim}} \mid T_i{=}1] \;-\; \mathbb{E}[Y_i^{\text{sim}} \mid T_i{=}0],
\]
and their difference
\[
\Delta=\widehat{\text{ATE}}_{\text{model}}-\widehat{\text{ATE}}_{\text{emp}}.
\]
We use $B=1000$ resamples and form percentile intervals from the empirical distributions. In the tables we therefore report (i) point estimates of $\widehat{\text{ATE}}_{\text{emp}}$ and $\widehat{\text{ATE}}_{\text{model}}$, (ii) their respective 95\% confidence intervals, and (iii) the signed discrepancy $\Delta$ with its own confidence intervals.

\subsection{Mobility allowance duration} 

The first exercise examines whether the model can replicate age-based differences in the duration of unemployment benefits, specifically the mobility allowance, a now-discontinued income support program for workers displaced due to firm closures in Italy. As shown in Figure \ref{fig:labour_force_status_composition}, less than 1\% of records in our global sample fall under this category. 

Under the policy rules in place until January 2017, the maximum benefit duration depended on the worker's age at the time of job loss. Workers over 40 were eligible for up to 24 months of benefits, while those under 40 could receive income support for up to 12 months \citep{brunello1997benefit}. This sharp policy threshold makes it an ideal setting to test whether the model internalises and reflects such causal discontinuities in allowance duration.

We drew individuals with at least one registered episode of mobility allowance from the test set who also have only standard employment or retirement spells in their histories. To increase homogeneity, we exclude individuals with any episode of self-employment, artistic or athletic careers, or voucher-based contracts. Additionally, we retain only those employed in northern Italian provinces at the time of firm closure, since a worker from the south was allowed one additional year of beneﬁts after 40 years of age. We restrict the sample to workers who began receiving the benefit between the ages of 30 and 50, and who have less than 1\% missing data on key variables: economic sector (ATECO), work province, income, and job title. The resulting sample consists of 1,502 individuals.  

For each individual, we provide the model with their life trajectory up to the year we first observe the receipt of unemployment benefits. The model then generates the sequence from that point onward. This setup allows us to test whether the simulated unemployment subsidy duration reflects the correct age-based discontinuity.

To quantify the effect, we exploit the policy cutoff at 40 years of age (480 months) and estimate a local average treatment effect (ATE) within symmetric windows $h \in \{12, 48, 96\}$ months around the threshold. Let $A_i$ denote age at first unemployment, and define the sample and treatment within the window, and treatment dummy $T_i$
\[
\mathcal{I}(h) = \{ i : |A_i - 480| \le h \}, \qquad T_i = \mathbb{I}[A_i \ge 480].
\]

Because an exogenous age cutoff determines the treatment, differences in unemployment duration just above and just below the threshold can be interpreted as a causal effect of the policy rule. Accordingly, the treatment effect is defined as a local difference in means, directly comparable between the empirical data and the model-generated sequences
\[
\widehat{\text{ATE}}(h) =
\mathbb{E}[Y_i \mid T_i{=}1,\, i \in \mathcal{I}(h)] -
\mathbb{E}[Y_i \mid T_i{=}0,\, i \in \mathcal{I}(h)],
\]
where $Y_i$ is the observed duration of the first unemployment spell. This corresponds to a simple difference in means above vs.\ below the cutoff inside the window, with bootstrap confidence intervals obtained by resampling individuals from $\mathcal{I}(h)$, and the signed discrepancy dependent on the window
\[
\Delta(h) = \widehat{\text{ATE}}_{\text{model}}(h) - \widehat{\text{ATE}}_{\text{emp}}(h).
\]


\subsection{Child Penalty} \label{mothers}

The second exercise evaluates the model’s capacity to reproduce the income trajectory experienced by women following childbirth, a well-established finding in the causal literature on gender inequality in the labour market \citep{kleven2019children, casarico2023behind}.  

From the test set, we selected a sample of women who had only held standard employment contracts throughout their recorded work histories. We identify mothers as those who experienced at least one maternity leave spell, defined by a maternity leave of at least two months in a given year. To ensure sample homogeneity, we exclude women who were ever classified as managers or professional athletes, as these careers may follow atypical trajectories. We retain only those with less than 1\% missing data across key variables (ATECO sector, work province, income, and job title). Finally, we restrict to women whose first childbirth occurs between the 5th and 20th observed year, ensuring sufficient pre- and post-treatment periods, and only mothers working in the southern regions of Italy. The resulting evaluation set contains 631 mothers.  

For each mother, we truncate the observed sequence at maternity leave ($t=0$) and generate forward from that point. In addition, we construct offset-1 variants where the sequence is truncated one year before the maternity leave ($t=-1$). In these offset-1 runs, the model is not forced to generate a maternity event in the sequence, providing a natural basis for counterfactual comparison.

Unlike pension and unemployment, where treatment is determined by exogenous or quasi-exogenous rules, fertility is a choice-driven event. To address this treatment endogeneity, we adopt an event study design that compares sharp changes in labour market outcomes around the birth of the first child for mothers (M) relative to non-mothers (NM). The timing of childbirth creates a discrete shock to mothers' careers, while other determinants of labour market outcomes would be expected to evolve smoothly over time. This approach also allows us to trace the full dynamic trajectory of the effects with precision, exploiting individual-level variation in the timing of first births \citep{kleven2019children}. 

For individual $i$ in year $s$ with event time $t$ (measured in years since first maternity leave) and treatment group $T \in \{\text{M, NM}\}$, we run the following regression separately for mothers and non-mothers:
\[
Y^{T}_{ist} = \sum_{j\neq -1} \alpha^{T}_j \mathbf{1}[t=j] \;+\; 
\sum_k \beta^{T}_k \mathbf{1}[\text{age}=k] \;+\; 
\sum_y \gamma^{T}_y \mathbf{1}[\text{year}=y] \;+\; \nu^{T}_{ist},
\]
omitting $t=-1$ as the reference period. This yields dynamic effects $\hat\alpha^{T}_t$ relative to the year before birth, while the flexible age and calendar-year fixed effects control for lifecycle and macroeconomic trends.  

To compute percentage penalties, we normalise with the counterfactual income path absent childbirth, defined by the predicted values from the age and year components,  
\[
\widetilde Y^{T}_{ist} = \sum_k \hat\beta^{T}_k \mathbf{1}[\text{age}=k] + \sum_y \hat\gamma_y \mathbf{1}[\text{year}=y], 
\qquad
\widehat{P}^{T}_t = \frac{\hat\alpha^{T}_t}{\mathbb{E}[\widetilde Y^{T}_{ist} \mid t]}.
\]

We construct two types of controls, i.e., non-mothers (NM), against which the post-birth path of mothers is compared:
\begin{enumerate}
    \item \emph{Matched non-mothers.} Using propensity scores based on pre-birth characteristics that correlated both with treatment assignment and outcomes (age, year of birth, area of birth, year start working, first sector), we match each mother to a non-mother on common support. The matched control is aligned to the mother’s event time so that $t=0$ corresponds to the mother’s first birth year.
    \item \emph{Simulated no-maternity trajectories.} For each true mother, we use the offset-1 simulations where the model does not generate any maternity leave episode. We then average them per mother-year to form a synthetic counterfactual trajectory aligned to the event time of the true.
\end{enumerate}

At event time $t$, we take the event-study outputs for mothers (treated, $\text{T}=\text{M}$) and the chosen control group (non-mothers or simulated no-maternity paths, $\text{T}=\text{NM}$) and define the percentage child penalty as the difference in the estimated event coefficients
\begin{equation}
  P_t \;=\; \widehat{P}^{(\text{T}=\text{M})}_t \;-\; \widehat{P}^{(\text{T}=\text{NM})}_t ,
\end{equation}
where each $\widehat{P}^{(\cdot)}_t$ is already reported in percentage units by the event-time regression.
We summarise the dynamic path by averaging over an event-time window
\begin{equation}
  P_{a\text{--}b}
  \;=\;
  \frac{1}{b-a+1}\sum_{t=a}^{b} P_t .
\end{equation}

For each pairing (empirical matching and model-based control) and each event time $t$ used in the window, we observe the estimated percentage effects and their standard errors for both arms:
\[
\Big(\widehat{P}^{(\text{T}=\text{M})}_t,\, \widehat{\sigma}^{(\text{T}=\text{M})}_t\Big)
\quad\text{and}\quad
\Big(\widehat{P}^{(\text{T}=\text{NM})}_t,\, \widehat{\sigma}^{(\text{T}=\text{NM})}_t\Big).
\]
We then implement a parametric (Gaussian) bootstrap:
\begin{enumerate}
  \item For every $t \in \{a,\ldots,b\}$, draw independently
  \begin{equation}
    \widetilde{P}^{(\text{T}=\text{M})}_t \sim \mathcal{N}\!\big(\widehat{P}^{(\text{T}=\text{M})}_t,\, (\widehat{\sigma}^{(\text{T}=\text{M})}_t)^2\big),
    \qquad
    \widetilde{P}^{(\text{T}=\text{NM})}_t \sim \mathcal{N}\!\big(\widehat{P}^{(\text{T}=\text{NM})}_t,\, (\widehat{\sigma}^{(\text{T}=\text{NM})}_t)^2\big).
  \end{equation}
  \item Form the bootstrap penalty path and its window average:
  \begin{equation}
    \widetilde{P}_t \;=\; \widetilde{P}^{(\text{T}=\text{M})}_t \;-\; \widetilde{P}^{(\text{T}=\text{NM})}_t,
    \qquad
    \widetilde{P}_{a\text{--}b} \;=\; \frac{1}{b-a+1}\sum_{t=a}^{b}\widetilde{P}_t .
  \end{equation}
  \item Repeat the previous steps $B$ times. The $95\%$ confidence interval for $P_{a\text{--}b}$ is the percentile interval over these runs.
\end{enumerate}

Let $P^{\text{model}}_{a\text{--}b}$ and $P^{\text{emp}}_{a\text{--}b}$ denote the window-averaged penalties computed from the model-based control and from the empirical matched control, respectively. We report the signed difference and its absolute value:
\begin{equation}
  \Delta_{a\text{--}b} \;=\; P^{\text{model}}_{a\text{--}b} - P^{\text{emp}}_{a\text{--}b}.
\end{equation}
A confidence interval for $\Delta_{a\text{--}b}$ is obtained from \emph{paired} parametric draws: on each bootstrap iteration $b$, we simulate
$\widetilde{P}^{\text{model}}_{a\text{--}b}$ and $\widetilde{P}^{\text{emp}}_{a\text{--}b}$ from their respective Gaussians (using the reported standard errors at each $t$) and record
$\widetilde{\Delta}_{a\text{--}b} = \widetilde{P}^{\text{model}}_{a\text{--}b} - \widetilde{P}^{\text{emp}}_{a\text{--}b}$. The percentile interval over these runs is reported as the $95\%$ confidence interval.

\subsection{Retirement Timing} 

As a final causal benchmarking exercise, we evaluate the model’s ability to replicate the well-documented finding that the month of birth influences retirement timing \citep{ardito2020work}. Specifically, individuals born earlier in the calendar year tend to retire slightly later, due to institutional and lifecycle timing mechanisms.

We select from the test set all male individuals born between 1940 and 1950, restricting the sample to those born in either January or December. To ensure a clean employment trajectory, we include only those who have been continuously and exclusively employed in standard employment, excluding individuals who were ever self-employed, para-subordinate, unemployed, or employed under special regimes (e.g., artists, athletes, or voucher-based contracts). We further restrict the sample to those whose first pension spell occurs between ages 55 and 68, and who exhibit no long gaps in their administrative records. To reduce noise from incomplete information, we also exclude individuals with missing values in key attributes such as work province, economic sector (ATECO), income, and job title. After applying these criteria, the final evaluation set comprises 1,010 individuals.

For each of these individuals, we truncate their observed sequence prior to retirement and let the model generate forward until 2015. We consider two cutoff horizons: one year before the observed retirement (1y) and four years before retirement (4y), with $K=8$ simulations per individual. The longer horizon increases predictive uncertainty but avoids anchoring the model too closely to the retirement event.

The empirical average treatment effect is the difference in mean retirement age (in months) between January- and December-born individuals in the evaluation set:
\[
\widehat{\text{ATE}} \;=\; 
\mathbb{E}[Y_i \mid T_i{=}1] \;-\; \mathbb{E}[Y_i \mid T_i{=}0],
\]
where $T_i=1$ if born in January and $T_i=0$ if born in December. In the pension application, treatment assignment is effectively exogenous, since the month of birth is orthogonal to individual career choices and labour market conditions. As a result, the treatment effect can be interpreted as a clean difference in means between the two groups of retirees.


\subsection{Out-of-sample Generation} \label{oos_gen}

We randomly selected 1,500 individuals from the test set, specifically drawn from those used in the three causal inference benchmarking exercises. For each individual, we generated their life trajectory for 20 years beyond the last year observed in the training data (2015).

The purpose of this experiment is to evaluate the structural coherence of out-of-sample generations over an extended horizon. For each generated sequence, we identify the first year in which the model produces a structural inconsistency. Consistency means that each calendar year must begin/end with an End-of-Year \texttt{<EOY>} token, followed by a valid combination of monthly tokens and corresponding durations. These can be empty, i.e., no event, or include a valid life event, but must respect the order: month, then optional content, then duration. For each year, there must be at least one event, even if empty (so no repeated \texttt{<EOY>} tokens). A duration token must be followed either by another month token or an \texttt{<EOY>} token indicating the start of the next year, and \texttt{<EOY>} tokens must always be followed by a month token.

Once a structural failure is detected in a given sequence, we record the year of failure and disregard any subsequent tokens, even if the sequence appears to recover. We then plot the density of the failure years to analyse how long the model maintains structural validity over time.

\section{Additional results}\label{additional results}

\begin{figure*}[!thp]
\includegraphics[width=.5\linewidth]{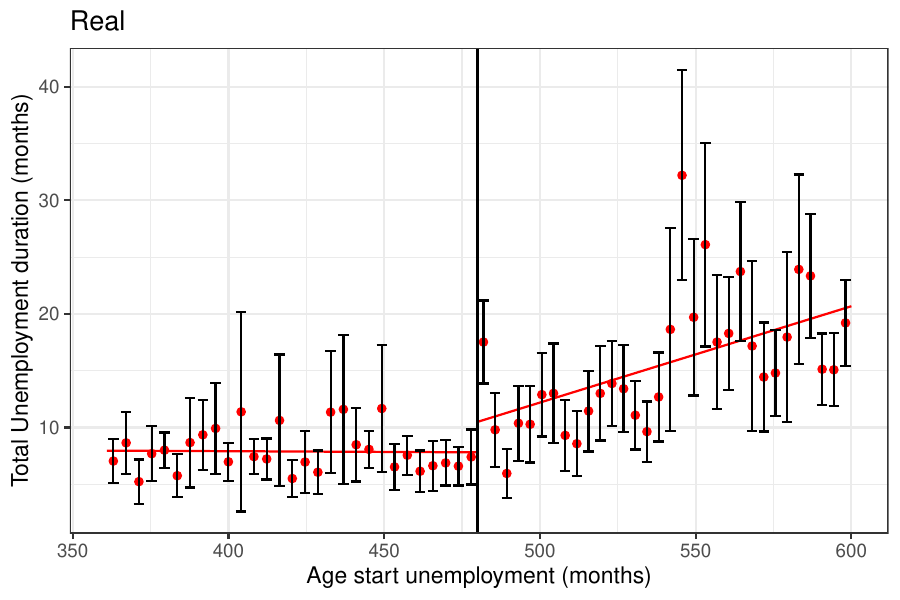}\hfill
\includegraphics[width=.5\linewidth]{figures/Generated_unemployment_decoder_only.pdf}
\caption{Real and model simulated average duration with 90\% confidence interval of mobility allowance duration as a function of worker age at job loss in months in a 10-year bandwidth around the 40-year-old threshold.}
\label{fig:unemployment_result_encoder_decoder_only}
\end{figure*}

Figure \ref{fig:unemployment_result_encoder_decoder_only} compares the real and simulated average duration of mobility allowance receipt as a function of age at displacement. Taken together, these results suggest that while the model does not reproduce the sharp discontinuity in unemployment benefit duration at the exact policy cutoff, it does capture the broader age gradient and approximates the average magnitude of the causal effect when evaluated in wider horizons around the treatment. This underscores both a limitation, i.e., the difficulty of learning from rare, rule-driven discontinuities, and a strength, i.e., when sufficient natural variation is present, the model’s simulated treatment effects align closely with their empirical counterparts.

\begin{figure}[!thp]
\centering
\includegraphics[width=.5\linewidth]{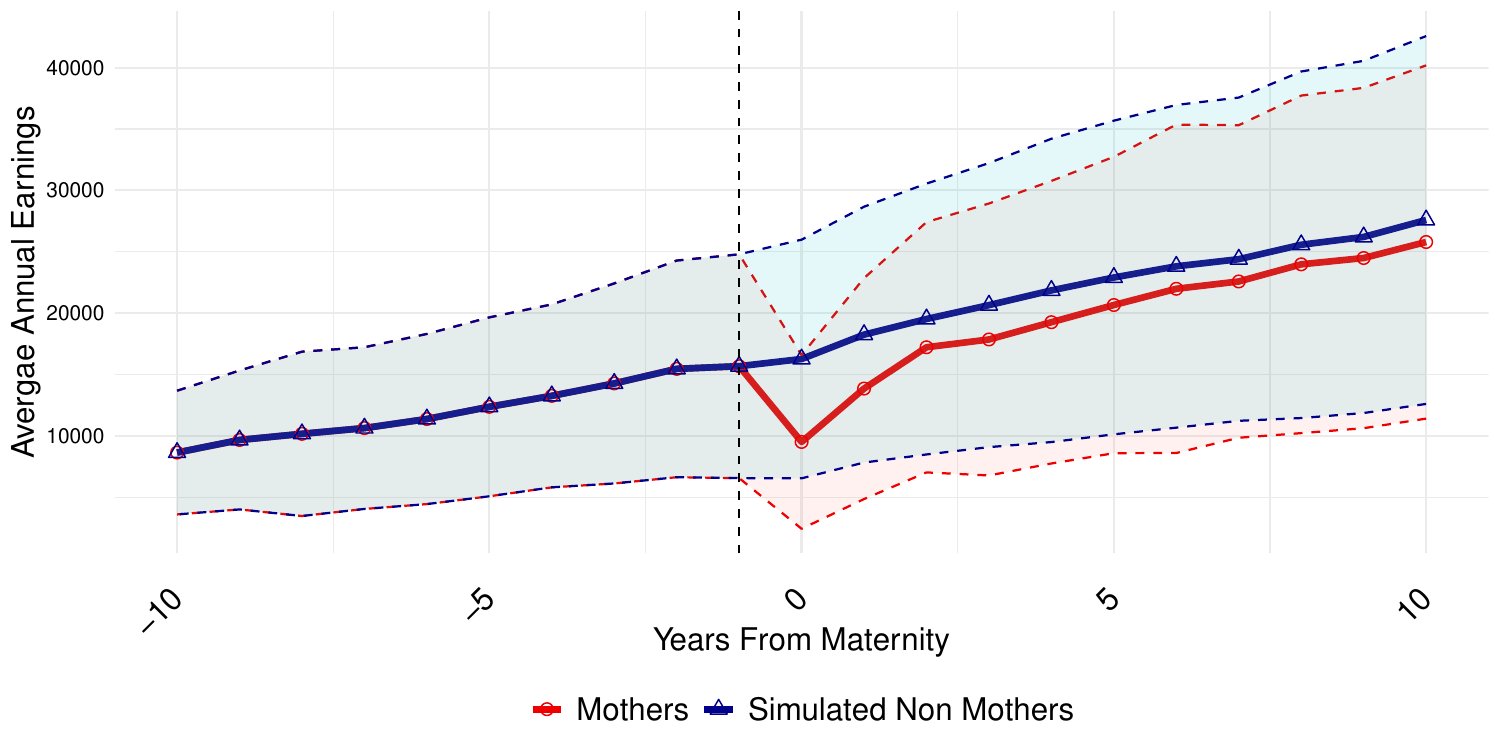}\hfill
\includegraphics[width=.5\linewidth]{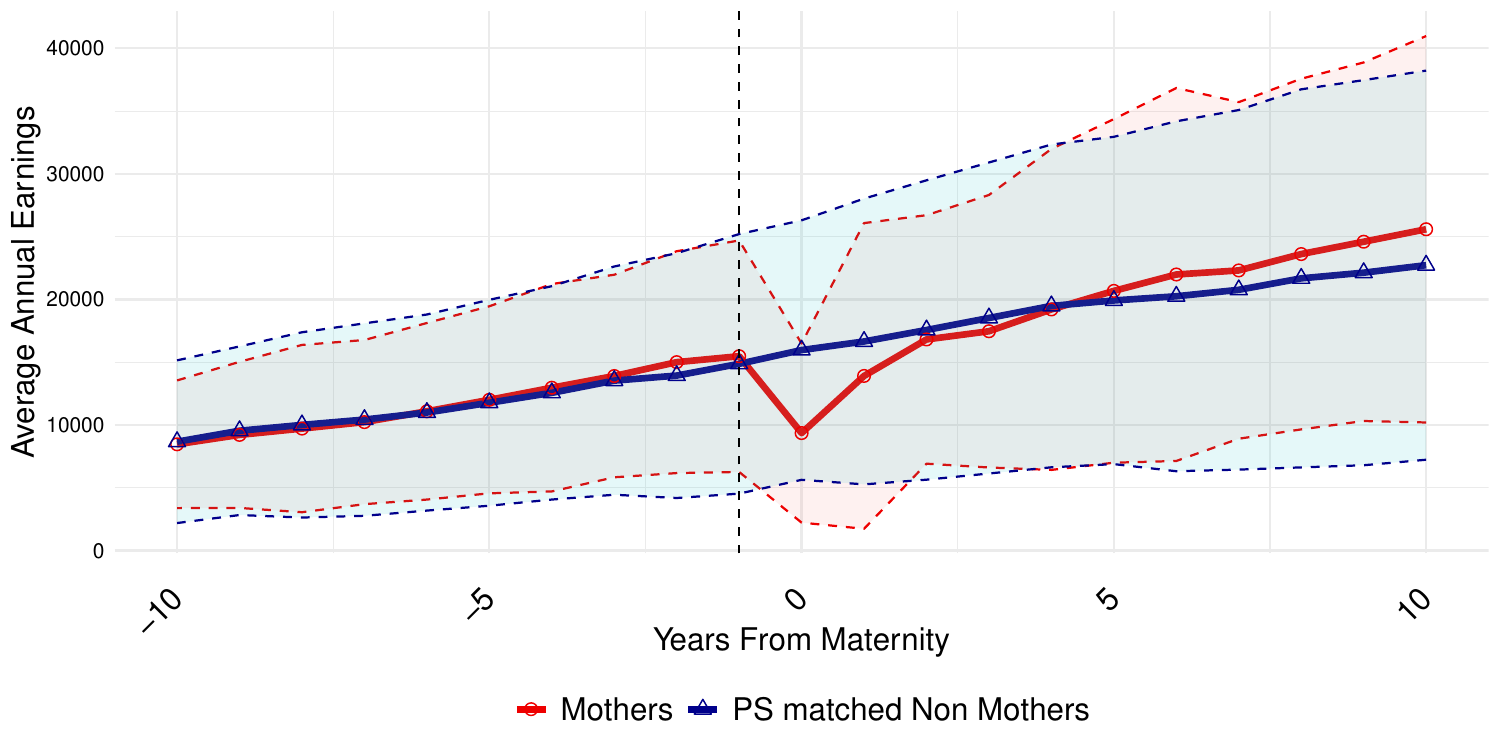}
\caption{Average annual earnings of mothers (circle dots) compared to two types of controls (triangle dots). The left panel uses model-generated controls based on offset--1 simulations without maternity events, and the right panel uses propensity-score–matched non-mothers. Shaded areas denote one standard deviation.}
\label{fig:mother_result_synth}
\end{figure}

Figure~\ref{fig:mother_result_synth} provides a first look at how mothers’ average earnings trajectories compare to two types of counterfactual controls: model-generated synthetic paths without maternity events (left) and propensity-score–matched non-mothers (right). Both approaches highlight the earnings divergence that emerges around childbirth, but they differ in persistence: the simulated control yields a child penalty that appears to persist over time, whereas with propensity-score matching, the point estimates attenuate toward zero by year 10 after birth. Notably, in the two years before maternity, we already observe a slight increase followed by a decrease in earnings, even before childbirth occurs, highlighting how difficult it is to pinpoint when treatment truly begins and what baseline trajectory should serve as the counterfactual. This contrast raises the broader question of what constitutes a good counterfactual group and appropriate conceptual questions: at what point does a woman \textit{become} a mother for the purpose of defining counterfactuals, and when are the relevant choices actually made, surely not at the precise year the first child arrives?

\begin{figure}[!thp]
\centering
\includegraphics[width=.5\linewidth]{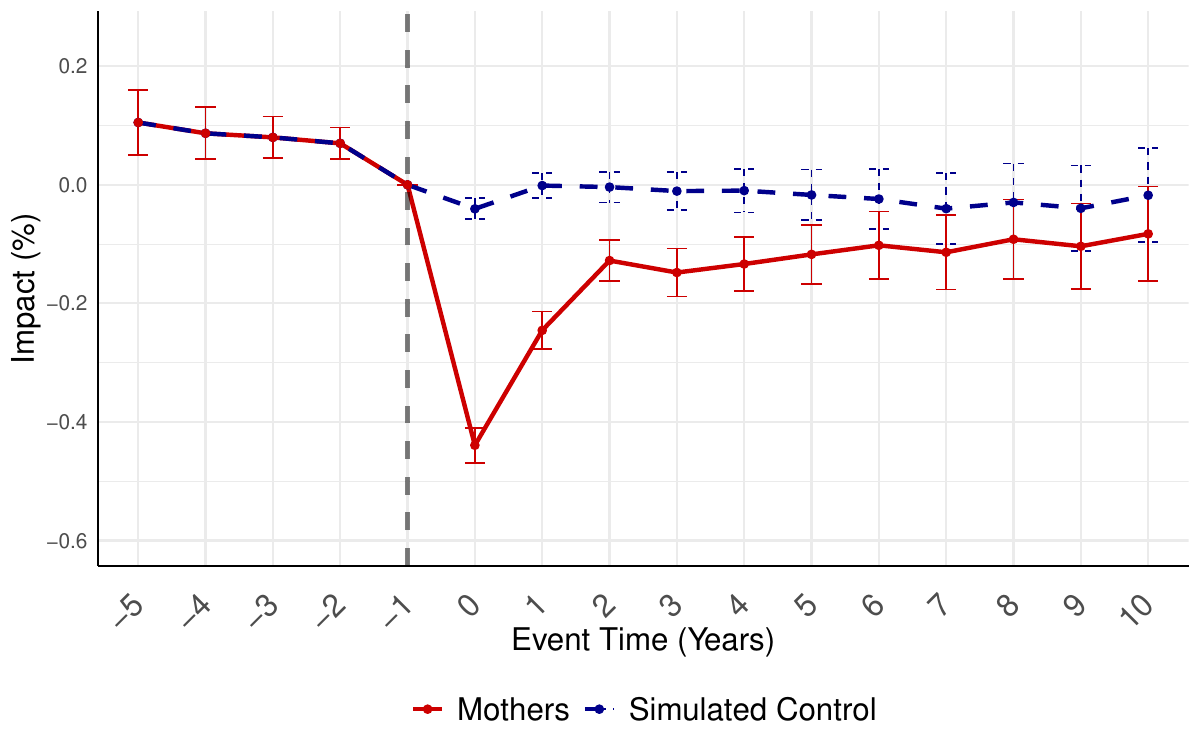}\hfill
\includegraphics[width=.5\linewidth]{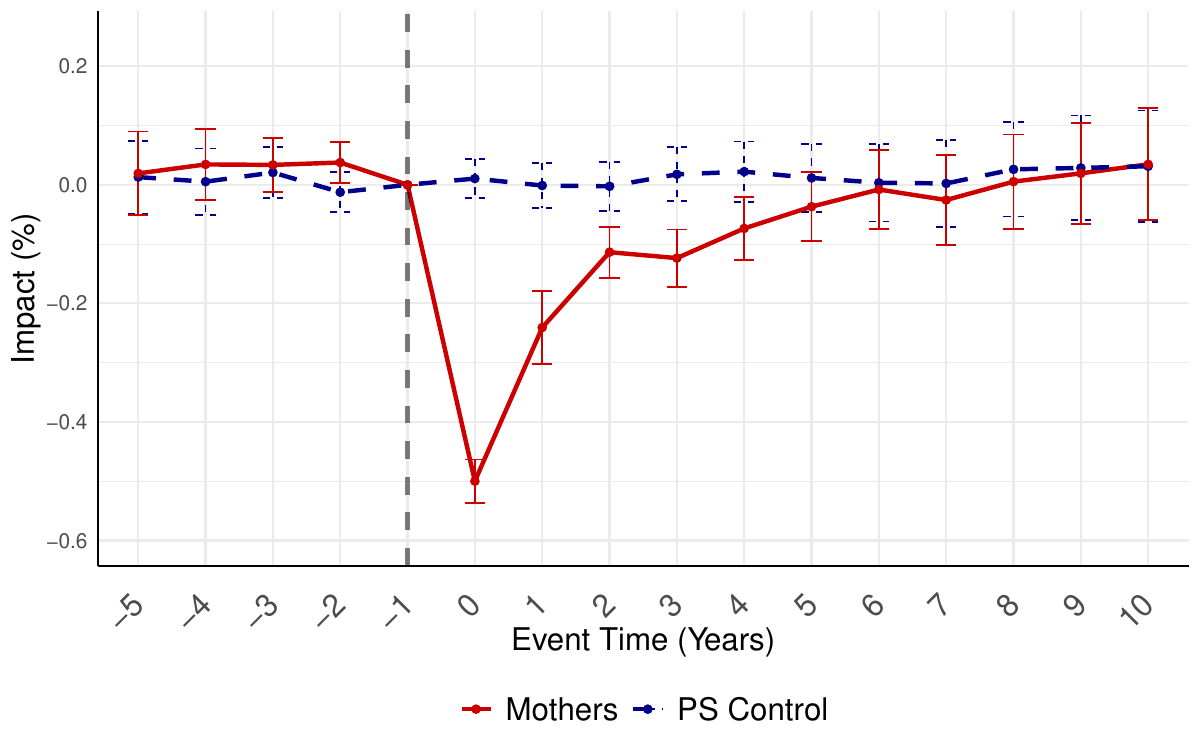}
\caption{Event time coefficients estimated as a percentage of the counterfactual
outcome $\widehat{P}^T_t$ relative the year before the maternity event ($t=-1$). The left panel uses model-generated synthetic controls based on offset--1 simulations without maternity events, while the right panel uses propensity-score–matched non-mothers as controls. The horizontal axis shows event time $t$ (years since birth), and vertical bars denote 95\% confidence intervals.}
\label{fig:mother_result_synth_Pt}
\end{figure}

Figure~\ref{fig:mother_result_synth_Pt} then summarises the event time coefficients estimated as a percentage of the counterfactual outcome absence childbirth, i.e., $\widehat{P}^T_t$, relative to the year before the maternity event ($t=-1$). The model-generated simulated controls (left panel) and propensity-score–matched non-mothers (right panel) yield broadly similar patterns, though with notable differences. In the simulated control case, the immediate short-run penalty tends to be smaller, but the estimates suggest a more persistent medium- to long-run effect. By contrast, the propensity-score comparison shows a sharper initial earnings drop and a quicker recovery. Taken together, these results show that the model reproduces empirical child penalties quite closely across horizons, while also revealing how alternative counterfactual constructions can shift the interpretation of the penalty’s timing and persistence.

\begin{landscape}
\begin{figure}[!thp]
\centering
\includegraphics[width=1\columnwidth]{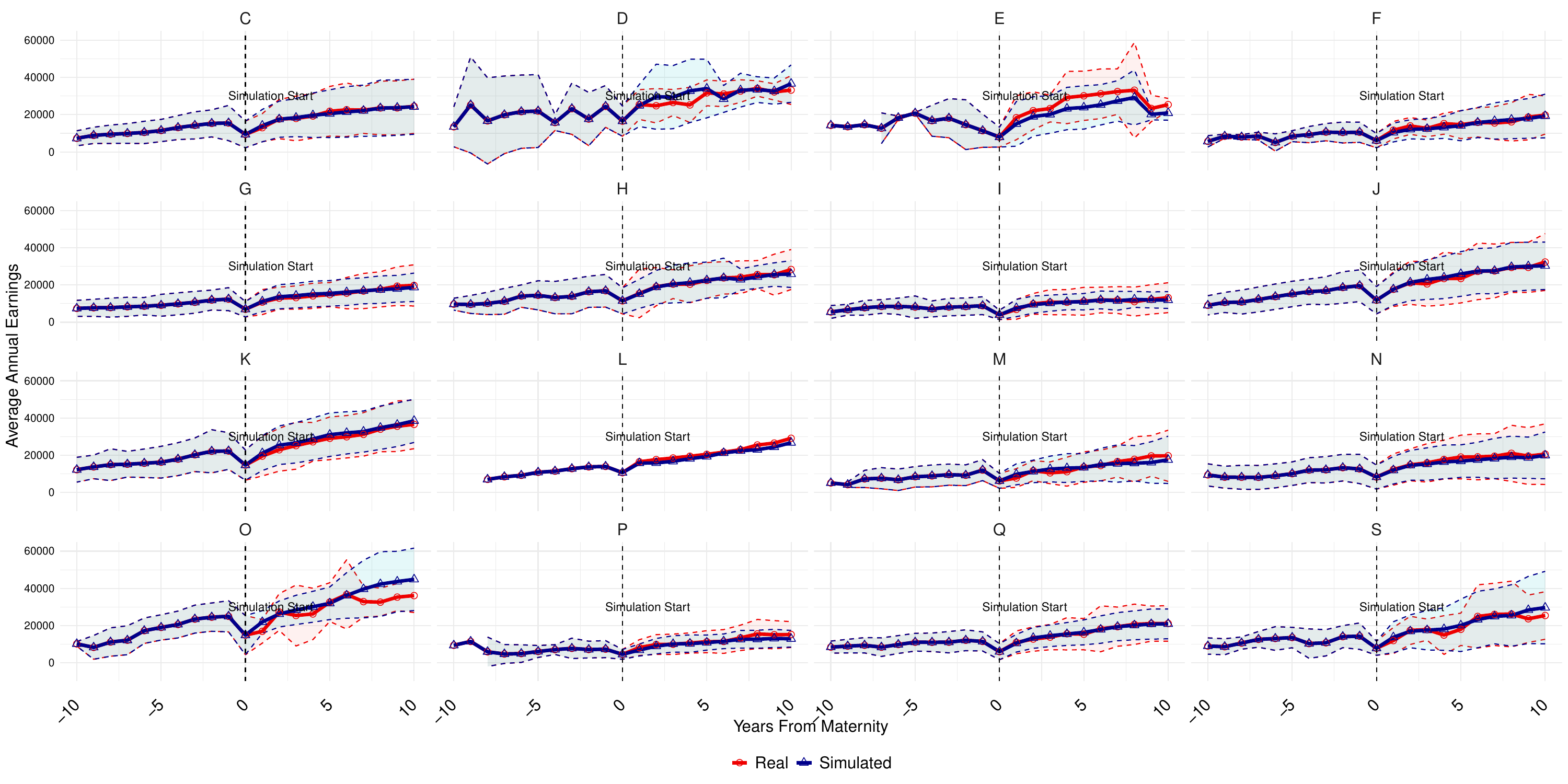}%
\caption{ATECO grouped real and model-generated trajectories of average annual earnings for working mothers in a 10-year window before and after their first maternity leave episode (Years From Maternity = 0). The solid lines represent the average monthly income, while the shaded areas denote the one standard deviation range.}
\label{fig:generated_mother_ATECO}
\end{figure}
\end{landscape}

\begin{figure}[!thp]
\centering
\includegraphics[width=.49\linewidth]{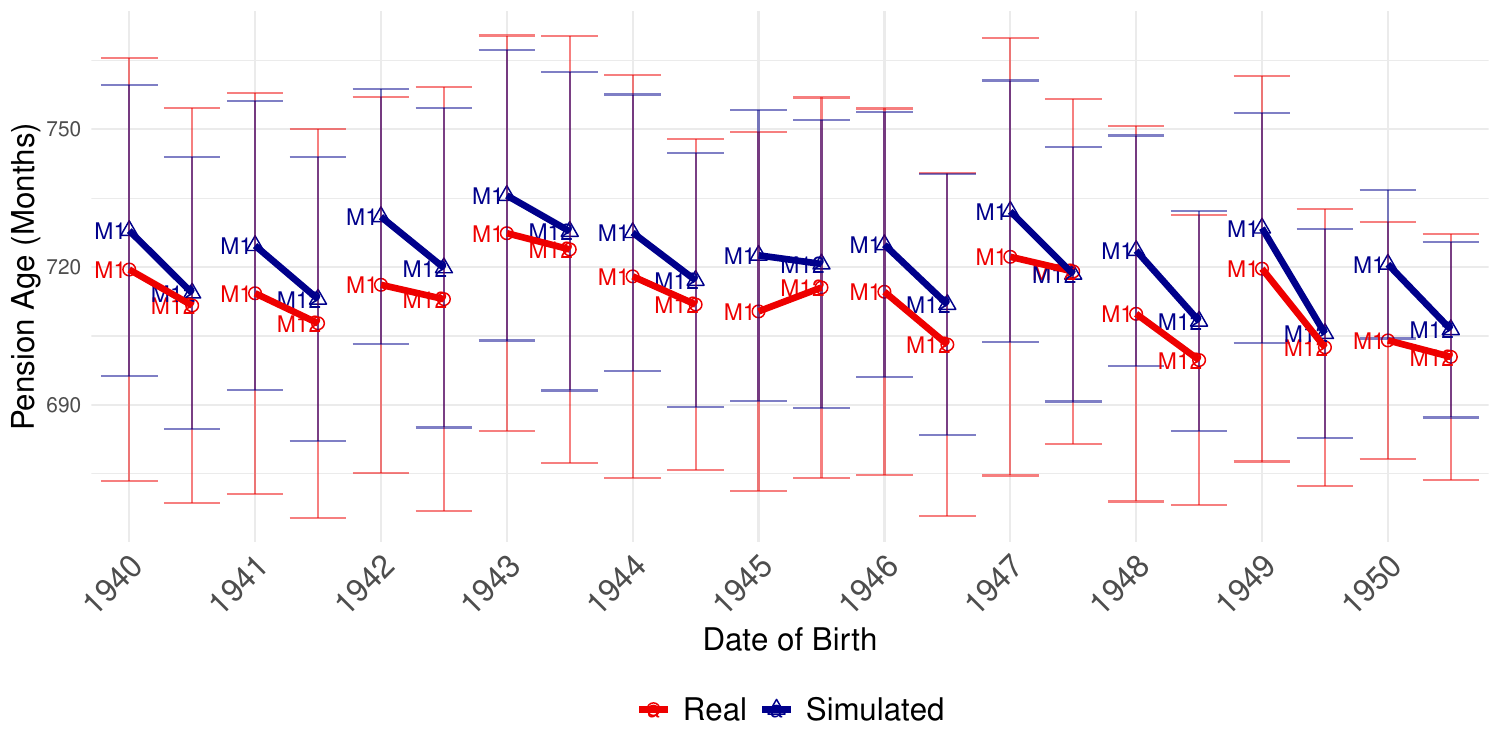}\hfill
\includegraphics[width=.49\linewidth]{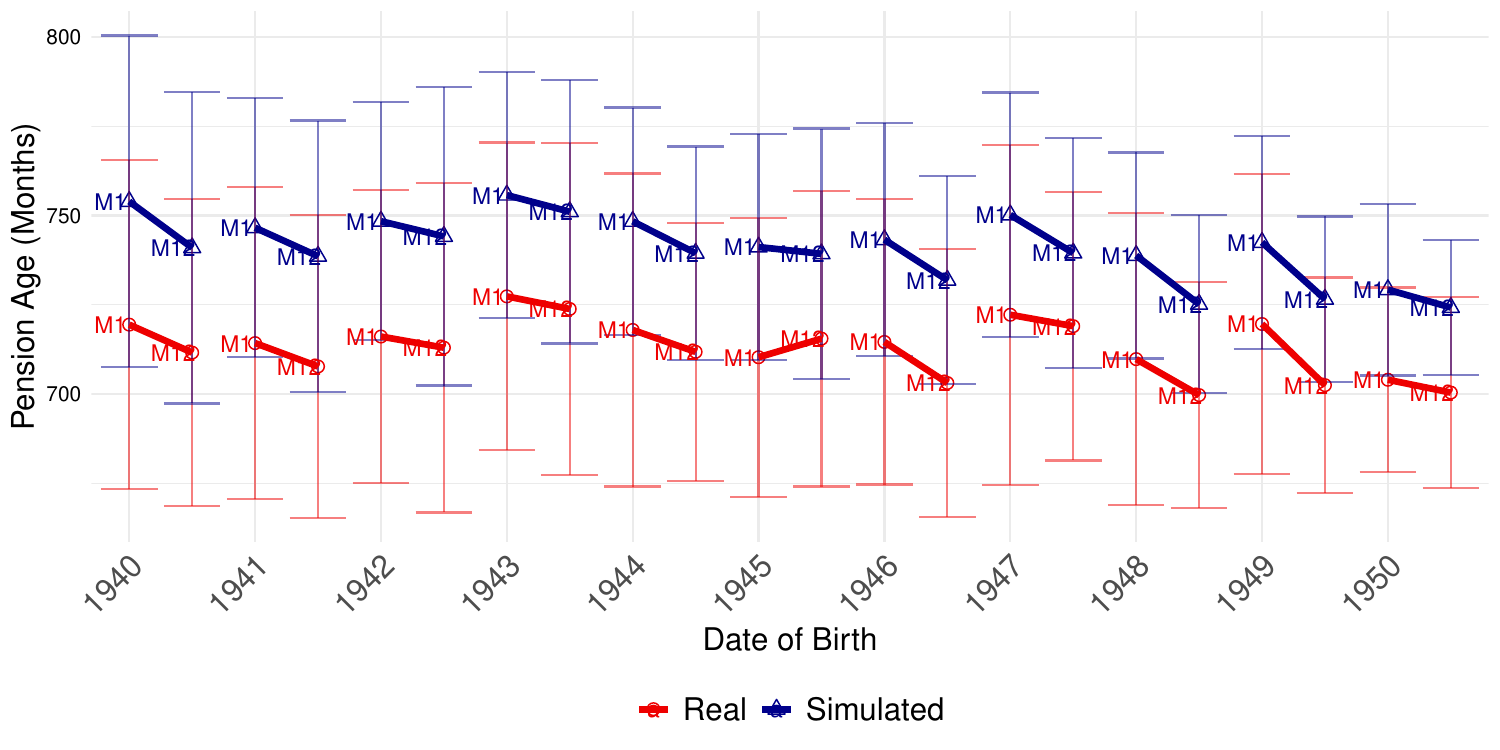}
\caption{Real and model-generated (blue) average retirement ages, measured in months, for male individuals born between 1940 and 1950. The left panel shows results using a cutoff four years before the observed retirement year, while the right panel shows results using a cutoff one year before. Each point indicates the cohort-specific average retirement age for January (M1) and December (M12), with vertical bars representing one standard deviation.}
\label{fig:pension_results_combined}
\end{figure}

Figure~\ref{fig:pension_results_combined} illustrates how the estimated retirement age varies with the choice of conditioning horizon. When the model is conditioned four years before retirement (left panel), it produces a clearer discontinuity between January- and December-born cohorts, whereas conditioning only one year before retirement (right panel) yields much smaller differences. This contrast underscores how sensitive the results are to where the observed sequence is cut: cutting further away forces the model to generate more of the trajectory, which amplifies the birth-month effect, while cutting closer reduces this reliance on simulated paths. These patterns highlight not only the model’s ability to reproduce well-documented empirical regularities but also the conceptual challenge of deciding how much information to condition on when constructing counterfactuals in retirement settings.

\begin{figure}[!thp]
\centering
\includegraphics[scale=0.45]{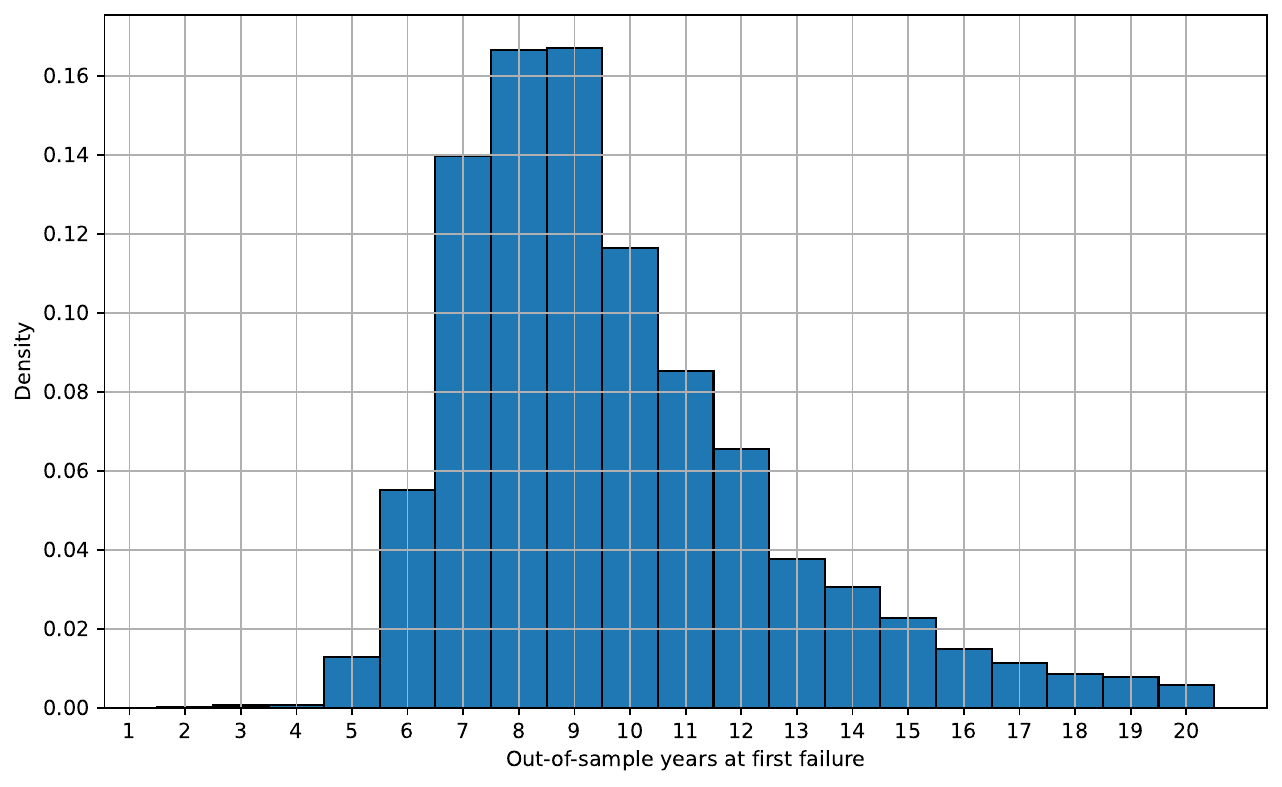}%
\caption{Density plot for the out-of-sample year when the model generates incorrect sequences, 5\% of the observations were able to complete the 20 years of generation without failures.}
\label{fig:out_of_sample_decoder_only}
\end{figure}



\begin{figure}[!thp]
\includegraphics[clip, trim=4.7cm 0cm 5.5cm 0cm, width=1\textwidth]{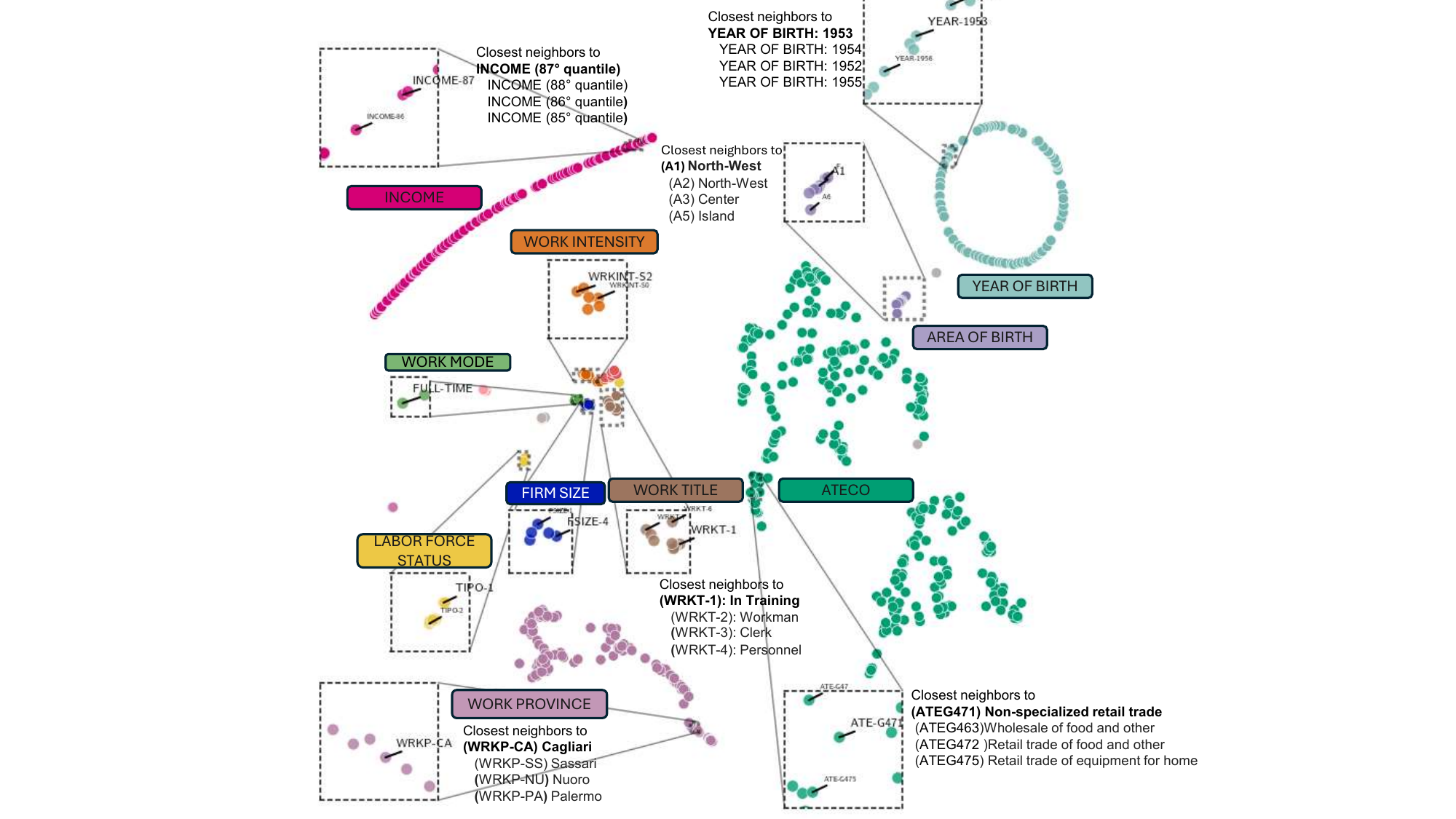}%
\caption{Two-dimensional projection of the learned embedding space from the model, visualised using the PaCMAP dimensionality reduction technique \citep{wang2021dimensionreductiontools}. Each point represents a token from the vocabulary ($\lvert \mathcal{V} \rvert = 661$), with different colours denoting different token categories. Several regions of interest are highlighted with zoomed-in views, each showing the top three nearest neighbours of a specific token. Neighbour relationships are computed using cosine distance in the original high-dimensional embedding space ($d = 240$).}
\label{fig:concept_space_Decoder_only}
\end{figure}

\end{document}